\documentclass[11pt,a4paper]{article}

\usepackage[utf8]{inputenc}

\usepackage[english]{babel}

\usepackage{hyperref}

\usepackage{comment}

\usepackage{bbm}
\usepackage{eurosym}
\usepackage{amsmath}

\usepackage{amsthm}

\usepackage{amsfonts}

\usepackage{graphicx}
\graphicspath{ {Figures/} }

\usepackage[sort,round]{natbib}

\usepackage[margin=1in]{geometry}

\usepackage{subfig}
 
\usepackage{booktabs}

\usepackage{setspace}
\usepackage{xcolor}

\newcommand{\re}{{\rm I\!R}}

\newcounter{ProblemCounter}
\renewcommand\theProblemCounter{\arabic{ProblemCounter}}

\newcounter{RemarkCounter}[section]
\renewcommand\theRemarkCounter{\arabic{RemarkCounter}}

\providecommand{\e}[1]{\ensuremath{ \text{E{#1}}}}

\title{Pricing sovereign contingent convertible debt}

\author{Andrea Consiglio~\footnote{
Corresponding author. University of Palermo,
Palermo, IT. andrea.consiglio@unipa.it} \\
Michele Tumminello~\footnote{
University of Palermo,
Palermo, IT. michele.tumminello@unipa.it} \\
Stavros A. Zenios~\footnote{University of Cyprus, Nicosia, CY and Wharton Financial Institutions Center, University of Pennsylvania, USA.
zenios.stavros@ucy.ac.cy}}

\date{First draft March 2017. This version for publication, November 2017}

\begin{document}

\maketitle

\vspace{-1cm}
\begin{center}
{\it 
Working Paper 16-05 \\
The Wharton Financial Institutions Center\\
The Wharton School, University of Pennsylvania, PA}
\end{center}

\begin{abstract}
We develop a pricing model for Sovereign Contingent Convertible bonds (S-CoCo) with payment standstills triggered by a sovereign's Credit Default Swap (CDS) spread. We model CDS spread regime switching, which is prevalent during crises, as a hidden Markov process, coupled with a mean-reverting stochastic process of spread levels under fixed regimes, in order to obtain S-CoCo prices through simulation. The paper uses the pricing model in a Longstaff-Schwartz American option pricing framework to compute future state contingent S-CoCo prices for risk management. Dual trigger pricing is also discussed using the idiosyncratic CDS spread for the sovereign debt together with a broad market index. Numerical results are reported using S-CoCo designs for Greece, Italy and Germany with both the pricing and contingent pricing models.     
\end{abstract}


\newpage
\tableofcontents

\vspace{1.0in}
{\small 
\section*{\small Acknowledgements}
An early draft of this paper was presented at the European Stability Mechanism, Bank of England, Bank of Canada, Federal Reserve Bank of Philadelphia, the World Finance Conference at New York, 6th International Conference of the Financial Engineering and Banking Society, the XI International Summer School on Risk Measurement and Control, and research seminars at Norwegian School of Economics and Stevens Institute of Technology, and benefited from the comments of numerous participants and from suggestions by Damiano Brigo, Rosella Castellano, Paolo Giudici, Mark Joy, Mark Kruger, Mark Walker.

\noindent
Stavros Zenios is holder of a Marie Sklodowska-Curie fellowship funded from the European Union Horizon 2020 research and innovation
programme under grant agreement No 655092.}
\newpage

\section{Introduction}\label{sec:Intro}

The Eurozone crisis and the record-breaking Greek sovereign default in particular (technically, a restructuring), highlighted the need for international legal procedures to deal with sovereign defaults. In September 2015 the United Nations General Assembly adopted a resolution on ``Basic Principles on Sovereign Debt Restructuring Processes''\footnote{Resolution A/69/L.84 at  \url{http://unctad.org/en/pages/newsdetails.aspx?OriginalVersionID=1074}}. With the debate on appropriate legal mechanisms ongoing, see, e.g., \cite{Li:2016}, proposals have also emerged for financial innovation solutions to the problem. Sovereign contingent convertible bonds (S-CoCo) with automatic debt payment rescheduling, have been suggested in academic and policy papers as a potential solution to sovereign debt crises  \citep{ConZen:2015b,Brooke:TOty0kZJ,BarEichMod:2012}. These papers advance several arguments on the merits of contingent debt for sovereigns which we do not repeat here. Our contribution was to make these proposals concrete by suggesting a payment standstill mechanism triggered when the sovereign's CDS spread exceeds a threshold, and to develop a risk optimization model demonstrating how contingent debt improves a country's debt risk profile. An alternative proposal are GDP-linked bonds with coupon payments linked to a country's GDP level or GDP growth, see, e.g. \cite{KamShil:2009,BorMau:2004,BOE:2015,ConZen:2018GDP}. These instruments are quite distinct from S-CoCo, and the pros and cons of each are discussed in \cite{BOE:2015}, highlighting the quest for financial innovation solutions to sovereign debt crises.  

The IMF recently published  a staff report with an extensive technical annex \cite{IMF-SCDI:2017,IMF-SCDI:2017annex} discussing broadly defined sovereign contingent debt instruments (SCDI) as a ``countercyclical and risk-sharing tool", which ``remain[s] appealing".  One of the three specific types of instruments are ``\textit{extendibles}, which push out the maturity of a bond if a pre-defined trigger is breached".

Our contributions are, first, to develop a pricing model for one type of extendibles, and, second, to develop state contingent pricing of these instruments for risk management. To achieve these objectives, we model a mean-reverting stochastic process  of CDS spreads. However, the risk factors underlying spread changes are time-dependent and shocks are persistent, and the risk models could break down during a crisis when they are most needed. To address this salient issue we develop models under regime switching, and this is a significant innovation of the paper.

We hasten to add that our contribution does not settle the debate on market-based vs institutional-based triggers, or the debate on extendibles vs GDP-linked bonds. However, it contributes to an understanding of the pricing of sovereign contingent debt, its risk profile, and how design parameters can affect prices and risks. 

Justification for using CDS spreads as the trigger is found in existing literature. An appropriate trigger must be accurate, timely and defined so that it can be implemented in a predictable way \citep{CalHer:2013}. CDS spreads qualify. More importantly, the trigger should be comprehensive in its valuation of the issuing entity, and current literature shows that the CDS market is becoming the main forum for credit risk price discovery. 

Having established CDS spreads as appropriate early indicators for credit risk, the question is then raised on how to model their dynamics. Investigations on what drives CDS spreads identify global changes in investor risk aversion, the reference country's macroeconomic fundamentals, and liquidity conditions in the CDS market \citep{BadCatJah:2013,FabGiaTsu:2016,Longstaff:2011}, but the relative importance of such factors changes over time \citep{HeiSun:2014}. \cite{AmaRem:2003} observe that yield spreads of corporate bonds tend to be many times wider than what would be implied by expected default losses alone ---a ``credit spread puzzle"---  so that research has been focusing on modeling CDS spread returns directly, instead of modeling their response to market fundamentals. This approach is advocated by \cite{ConKan:2011} who provide modeling guidance by analyzing stylized facts of corporate CDS spreads and spread returns. Their  work identified important properties of the dynamics of CDS
spread returns --- stationarity, positive auto-correlations, and two-sided
heavy tailed distributions--- and they proposed a heavy-tailed multivariate time series model to reproduce the stylized properties. \cite{BriAlf:2005} develop a  shifted square-root diffusion model for interest rate and credit derivatives, and \cite{Donoghue:2014} develop a one-factor tractable stochastic model of spread-returns with mean-reversion (SRMR)
as an extension of Orstein-Uhlenbeck process with jumps.

These models were developed for corporate CDS but in principle they could be used for sovereign CDS as well. However, there is a prevalent issue with {\em regime switching} in the sovereign market, especially during crises. This became apparent to us while calibrating the SRMR model to Greek sovereign CDS spread data for our earlier paper. Calibration was unsuccessful for the period December~2007--February~2012, but converged when applied to different regimes identified using the test of \cite{BaiPerron:1998}. Therefore, we develop the regime switching mechanism instead of the jump process, and maintain the mean-reversion one-factor model of spread returns within each regime.

Regime switching in CDS spreads has been studied systematically by others as an empirical feature of the market, but, to the best of our knowledge, did not receive any attention in CDS pricing literature. \cite{FonSch:2010} find that euro area credit markets witnessed significant repricing of credit risk in several phases since 2007. They find a structural break in market pricing, which coincides with the sharp increase in trading of CDS and declining risk appetite of investors since summer 2007, and attribute these changes to flight-to-liquidity,  flight-to-safety, and limits to arbitrage. Regime switching in the corporate CDS market was identified by \cite{ConKan:2011} who find the behavior of spreads  ``clearly divided into two regimes: before and after the onset of the subprime crisis in 2007". These observations are consistent with the analysis of \cite{Augustin:2014JOIM} who finds that CDS spreads change abruptly in response to major financial events, such as, for instance, the Bear Stearns bailout and Lehman Brothers bankruptcy, and are very persistent otherwise, over a sample of 38 countries in the period 9 May 2003--19 August 2010. \citet{AleKae:2008} examine the empirical influence of a broad set of determinants of CDS spreads listed in iTraxx Europe, and find that, while most theoretical variables do contribute to the explanation of spread changes, their influence depends on market conditions. CDS spreads may behave differently during volatile periods compared to their behavior in tranquil periods. Using a Markov switching model they find evidence supporting the hypothesis that determinants of credit spreads are regime specific. \cite{CasSca:2014} find that, for corporate CDS, it is the volatility of returns that carries the signal, and they model regime switching using a hidden Markov matrix. Not only there is ample empirical evidence of regime switching, there are also theoretical arguments to support the observations. \cite{ArgKont:2016} use earlier models by Krugman and Obstfeld to argue that Greece can be in one of three regimes: one with credible commitment to stay in the eurozone with guarantees of fiscal liabilities, one that guarantees fiscal liabilities for as long the country stays in the eurozone but uncertainty about the country's commitment to do so, and one without fully credible commitment to the eurozone.    

Regime switching is a salient feature for our work because of the payment standstill triggered in case of a crisis, and crises typically signal a regime switch. For instance, during the eurozone crisis, a sharp drop of CDS spreads was noted across the board in the second half of 2012 following the ECB OMT announcement, and this was primarily due to a switch of the investors' sentiment, while country specific fundamentals remained broadly unchanged \citep{HeiSun:2014}. Hence, we develop our model with regime switching.

The rest of the paper develops the pricing model, and uses it to develop state-dependent prices at some risk horizon and simulate holding period returns. We start in Section~\ref{s:facts} with a statistical analysis of CDS spreads and spread returns for sovereigns in the eurozone periphery and core countries, and identify regime switching. This section informs our modeling work by giving a descriptive analysis of the eurozone sovereign CDS market. Section~\ref{sec:scengen} develops the scenario generating stochastic processes for both regime switching and steady state for CDS spreads, spread returns and risk free rates. Section~\ref{s:modeling} develops the pricing model, state-contingent pricing, and holding period return scenarios. We illustrate numerically for a eurozone crisis country (Greece) and core countries (Germany and Italy). Section~\ref{s:conclusions} concludes. The asymptotic modeling of CDS spreads ---as opposed to spread returns addressed in existing literature--- is given in  Appendix~\ref{app:asymptotic}.

\section{Some observations on sovereign CDS spreads} \label{s:facts}

A pricing model should be guided by the stylized facts of the observed series. The simulation window for pricing S-CoCo is 20 to 30 years, and the risk horizon for state contingent S-CoCo pricing is 10 to 20 years, so we focus on long term characteristics of the data generating process. We model and calibrate the limiting dynamics of spreads (Appendix \ref{app:asymptotic}), so we need the statistics describing time-dependent equilibria of the process. These equilibria are the {\em regimes}. 

\begin{figure}
\centering
\begin{tabular}{c}
\subfloat[Germany.]{\includegraphics[scale=0.25]{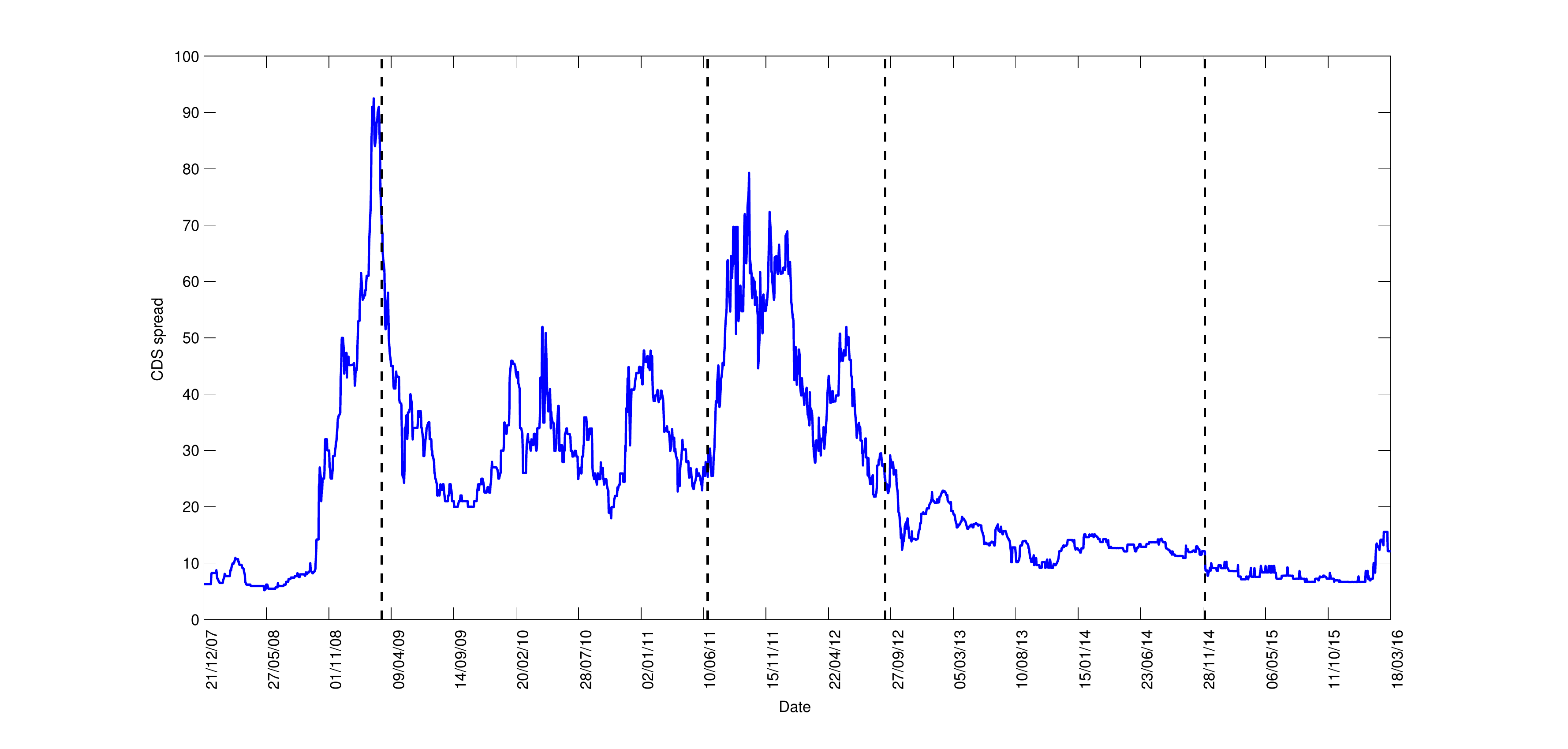}}\\
\subfloat[Italy.]{\includegraphics[scale=0.25]{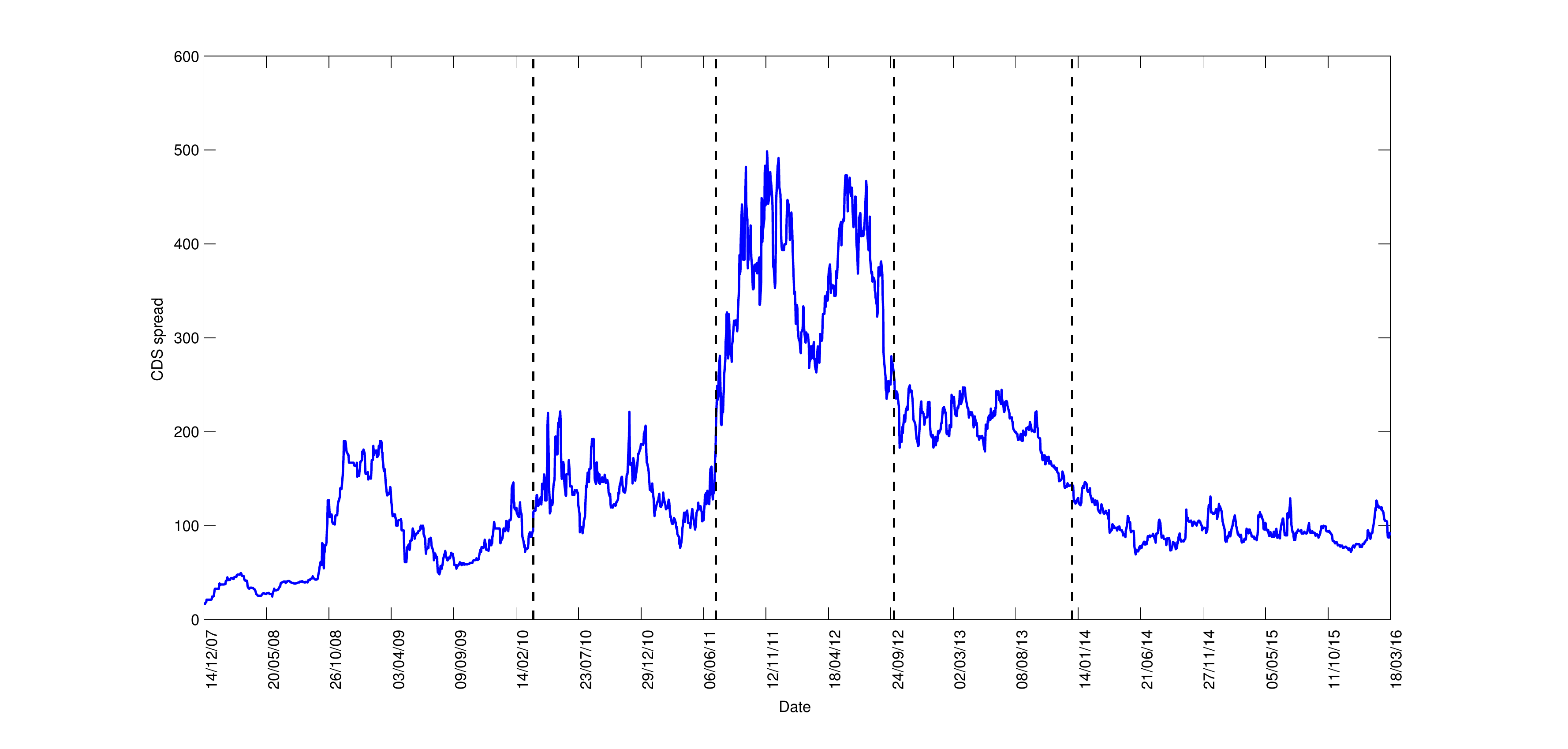}} 
\\
\\
\subfloat[Greece.]{\includegraphics[scale=0.25]{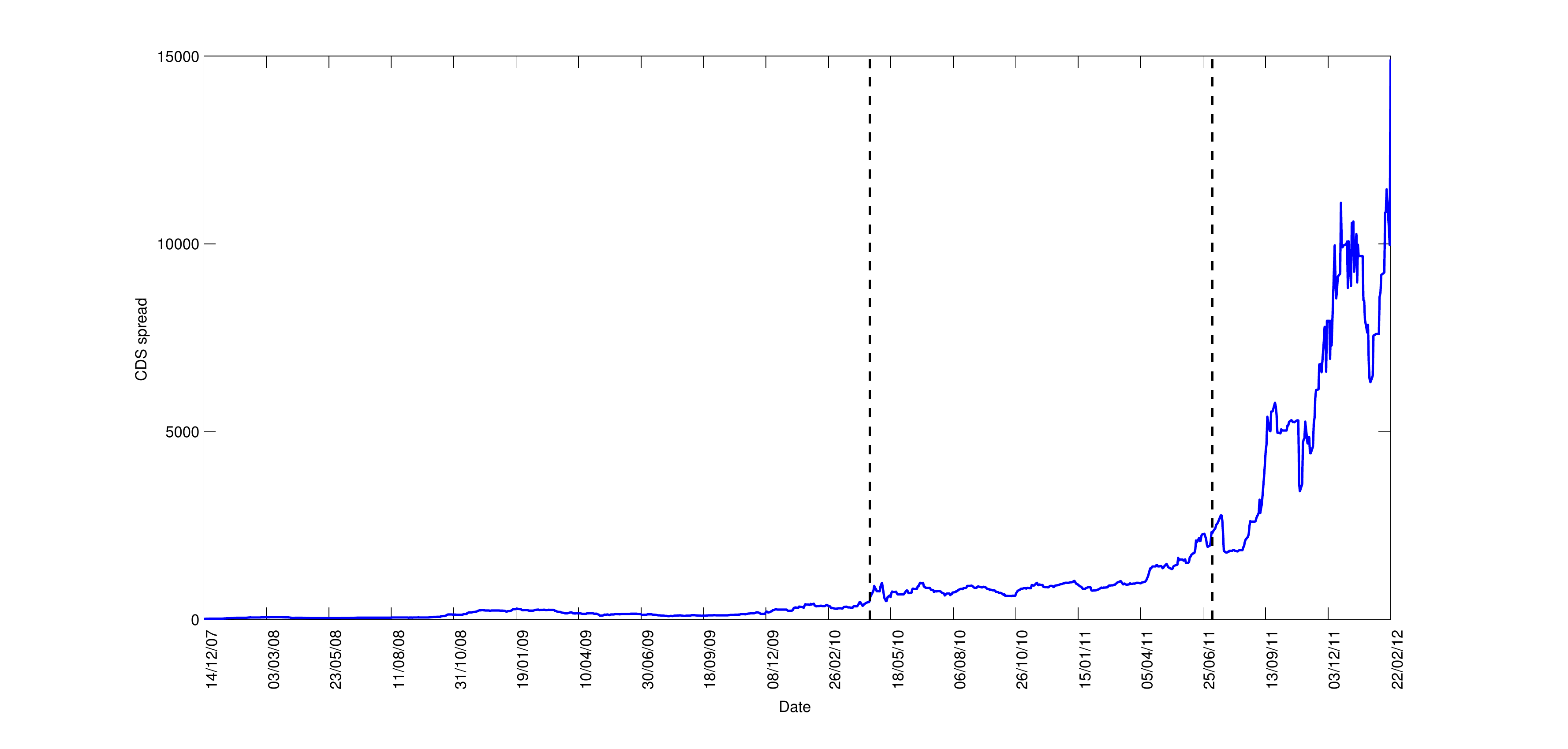}}
\end{tabular}
\caption{Regime switching identified using Bai-Perron test for the countries used to test the pricing models.}
\label{fig:B-P-tests}
\end{figure}

\begin{figure}
\includegraphics[width=\textwidth,height=0.5\textwidth,clip]{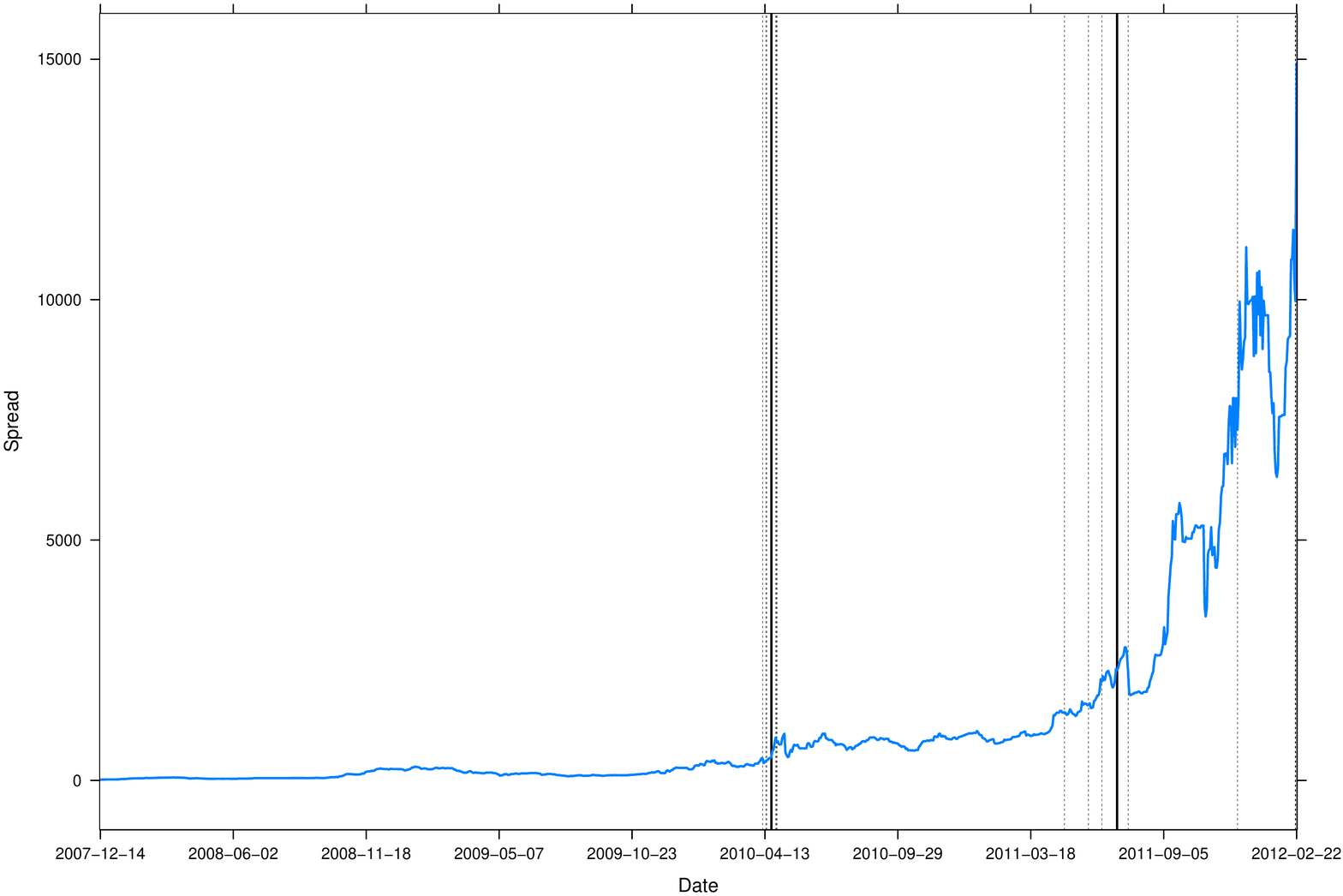}
\caption{Time series of the 5-yr CDS for Greece. Vertical dotted lines denote events which affect CDS spreads as given in Table~\ref{tab:Greece-events}. These events usher in new regimes that can be identified using using Bai-Perron test and denoted by the vertical solid lines.}\label{fig:Greece}
\end{figure}

In \cite{ConLotZen:2017} we analyzed the 5-yr CDS spread for a sample of European countries using daily data from February 2007 to March 2016. The test of \cite{BaiPerron:1998} applied to  the spread level identifies regime changes for all countries in the sample\footnote{We use the Bai-Perron test in the free software system R.}. Some countries, such as France, Italy, Portugal, Spain and Cyprus, are synchronized in their regime switching, whereas Germany, Ireland and Greece have idiosyncratic regime changes. For instance, only Germany had a regime switch associate with the subprime crisis and the collapse of Lehman Brothers in September 2008, while the onset of the eurozone crisis in spring 2010 signals regime switching for all countries. Ireland and Greece had their own idiosyncratic banking and sovereign debt crises, respectively, which ushered in new regimes. Figure~\ref{fig:B-P-tests} illustrates the CDS spreads and identifies regime changes for the three countries we will be using to test our pricing models. In particular, Germany with very low spread levels and low volatility, Greece with excessive debt undergoing a major crisis, and Italy with high debt levels and medium CDS spreads. Figure \ref{fig:Greece} displays the 5-yr CDS spreads for Greece, highlighting the major events that impact spreads, as summarized in Table~\ref{tab:Greece-events}. April 2010 signals switching from a {\em tranquil} to a {\em turbulent} regime of the Greek economy, and the events clustered around the change of regime are given in the table. The change of regime in July 2011 is the run up to the Greek PSI signaling the start of the Greek debt crisis. The events highlighted involve an open letter to European and international authorities by German finance minister Sch{\"a}uble about ``fair burden sharing between taxpayers and private investors" in providing financial support to Greece, and Jean-Claude Juncker's backing Germany's proposal arguing for ``soft debt  restructuring" with private sector participation.

\begin{table}
\centering
\begin{tabular}{l l}\hline
25/02/10 & EU and IMF mission in Athens delivers grim assessment of country's finances \\ 
16/03/10 & Eurozone finance ministers agree to help Greece but reveal no details \\
19/03/10 &  Prime Minister Papandreou warns Greece may have to go to the IMF \\
22/03/10 & President Barroso urges member states to agree aid package for Greece \\
12/04/10 & Greece announces that first trimester deficit was reduced by 39,2\% \\
13/04/10 & EU leaders agree bailout plan for Greece \\
14/04/10 &  ECB voices its support for the rescue plan of Greece \\ \hline
\end{tabular}
\caption{Major events relating to the Greek sovereign crisis regime switch of July 2011.} \label{tab:Greece-events}
\end{table}

The mean and standard deviation of CDS spreads and spread returns for different regimes are in Table~\ref{tab:meanCDS} for Germany, Greece, and Italy. These quantities are needed to calibrate the simulation model. We focus on regime switching for the spreads, but regime breaks can also be identified for the volatilities as suggested by \cite{CasSca:2014}. A comprehensive empirical analysis of the sovereign CDS markets, its statistical properties,  and regime switching analysis including regime switching identification with common regimes, is reported in \cite{ConLotZen:2017}.

\begin{table}
\centering
\begin{tabular}{l c c c c } 
{\bf Country} & {\bf Regime} & {\bf Spread} & {\bf Spread} & {\bf Spread return} \\
              &              & {\bf mean}       & {\bf std. dev.} & {\bf std. dev.} \\ \hline  \hline 
Germany   
&12/21/07--03/13/09   &22.22 & 23.54 & 5.48 \\
&03/16/09--06/20/11   &31.69 &  8.85 & 5.48\\ 
&06/21/11--09/12/12   &45.60 & 13.98 & 6.24\\
&09/13/12--12/02/14   &14.64 &  3.73 & 3.95\\ 
&12/03/14--03/18/16   & 8.25 &  1.93 & 6.31\\ \hline  
Greece    
& 12/14/07--04/20/10   & 146.09 & 103.90 & 4.45 \\
& 04/21/10--07/06/11   & 980.27 & 363.36 & 5.20 \\
& 07/07/11--02/22/12   &5770.43 &2917.45 & 8.05 \\ \hline 
Italy    
& 12/14/07--03/29/10   & 79.71 & 45.48 & 5.01 \\
& 03/30/10--07/07/11   &137.69 & 28.54 & 6.51\\
& 07/08/11--10/02/12   &361.94 & 68.99 & 4.94\\
& 10/03/12--12/27/13   &203.73 & 26.66 & 2.91\\
& 12/30/13--03/18/16   & 97.31 & 15.82 & 3.67\\ \hline 
\end{tabular}
\caption{CDS spread and spread return statistics in each one of the regimes identified using Bai-Perron test.}\label{tab:meanCDS}
\end{table}

\section{Scenario generating process} \label{sec:scengen}

Our scenario generator consists of a core process which determines regimes of the expected value of the CDS spread, and a 
process of the dynamics of the CDS spread superimposed on the mean value in each regime. In the next two subsections we model these sub-processes.

\subsection{Regime switching process}

We assume that regime transitions are driven by a discrete time-homogeneous Markov chain with finite state space $\mathcal{R} = \left \{ 1, 2, \ldots, S \right \}$, where
\begin{equation*}
p_{ij} = \mathbb{P}( X_k = j | X_{k-1} = i )
\end{equation*} 
is the transition probability of switching from regime $i$ at time $k-1$ to regime $j$ at time $k$. The transition probabilities matrix $P=\{p_{ij}\}$ is a stochastic matrix, i.e., $p_{ij} \geq 0$, for all $i,j \in \mathcal{R}$, and $\sum_{j \in R} p_{ij} = 1$, for all $i \in \mathcal{R}$.

The transition matrix $P$ is fundamental to simulating a regime switching process.
However, it cannot be estimated from observed historical series because regime breaks are rare events. Instead we infer $P$ from an estimate of the limiting probability $\pi^*$ (see definition below). We denote by $\pi_i^{(k)}$, for all $i \in \mathcal{R}$, the distribution at time $k$ of a Markov chain $X$,
\[\pi_i^{(k)} = \mathbb{P}(X_k = i). \]
Given a transition matrix $P$, it is possible to show that
\begin{equation}
\pi_j^{(k)} = \mathbb{P}(X_k = j) = \sum_{i \in \mathcal{R}} \mathbb{P}(X_k = j|X_{k-1} = i) \mathbb{P}(X_{k-1} = i) =\sum_{i \in \mathcal{R}} p_{ij} \pi^{(k-1)}_{i}\label{eq:DistributionDefinition}.
\end{equation}
If we denote by $\pi^{(k)}$ the row vector of probabilities $(\pi_1^{(k)},\cdots,\pi_S^{(k)})$, then (\ref{eq:DistributionDefinition}) is written in matrix form as 
\[
\pi^{(k)} = \pi^{(k-1)} P.
\]
\noindent 
Row vector of probabilities $\pi^{*}$ is a \textit{stationary distribution} for the Markov chain $X_k$, $k > 0$, if
\[\pi^* = \pi^* P, \mbox{ i.e., } \pi_j^* = \sum_{i \in \mathcal{R}} \pi^*_i p_{ij}.\]
Note that $\pi^*$ does not necessarily exist, nor it is unique. If $\pi^*$ exists and is unique then we can interpret $\pi_i^*$ as 
the average proportion of time spent by the chain $X$ in state $i$.

Given $P$, the stationary probability distribution $\pi^*$ is obtained as the solution, if it exists, of the following system:
\begin{align}
\pi^* &= \pi^* P\\
\pi^* \,\mathbf{1} &= 1\\
\pi^* &\geq 0.
\end{align}    

We assume that the stationary distribution can be estimated by the average 
number of days the CDS spread process is in regime $i$,
\begin{equation}
\hat{\pi}_i^*= \frac{\mbox{\textrm{Number of days CDS spread is in regime $i$}}}{\mbox{\textrm{Number of total days in sample}}}. \label{eq:EmpiricalSteadyState}
\end{equation}
This is a reasonable assumption for long horizons, but any estimate of the probability of a country being in a given regime can be used as well. For instance, we can use the transition probabilities of the rating agencies to estimate the likelihood of a country migrating to a better or worse regime from where it is at present. Each rating class implies a probability of sovereign default and, consequently, a CDS regime, so that the migration probabilities provide an estimate of the stationary distribution. In Section~\ref{sec:CoCoPricing} we carry out sensitivity analysis on the impact of these estimates on S-CoCo prices.
 
A constraint set on $P$ is obtained by the properties of square matrices from linear algebra theory. In particular, let us assume that the Markov matrix $P=\left ( p_{ij} \right ) \in \mathbb{R}^{S \times S}$  
has $S$ distinct eigenvalues denoted by $\lambda = \left [ \lambda_1 \lambda_2 \ldots \lambda_S\right]$\footnote{We are using matrix diagonalization, and the same conclusions are obtained when eigenvalues are not distinct, but the corresponding eigenvectors are linearly independent. Since we can arbitrarily choose to have a transition matrix with distinct eigenvalues, we present our analysis only for this case.}. Since $P$ is a stochastic matrix, the eigenvalue with highest magnitude has absolute value equal to one, $|\lambda_1| =1$, and according to the Perron-Frobenius theorem $1 = \lambda_1 > |\lambda_i|$, for all $i=2,3, \ldots, S$. 
Denote by $\xi_i$ the row vector which is the left eigenvector associated with the eigenvalue $\lambda_i$ of $P$, and denote by $\nu_i$ the column vector which is the right eigenvector of the same $\lambda_i$, with $\xi_i$ and $\nu_i$ obtained by solving
\begin{align}
\xi_i P &= \lambda_i \xi_i\\
P \nu_i &= \lambda_i \nu_i.
\end{align}
Note that the left and right eigenvectors are orthonormal, so $\xi_i \cdot \nu_j = \delta_{ij}$, where $\delta_{ij}$ is the Kronecker delta. 

Also observe that the right eigenvector for  $\lambda_1=1$ is a unit vector as $P$ is a stochastic matrix and all the rows sum up to 1, i.e., 
\[
P \nu_1 = \nu_1.
\]
Furthermore, if $P$ is the transition matrix of a stationary process, then the left eigenvector for $\lambda_1$ is the steady distribution $\xi_1 = \pi^*$, and we have
\[
\xi_1 P = \xi_1.
\]
Denote by $U= \left ( u_{ij} \right)$ a matrix whose columns are the right eigenvectors of $P$, and by $V=\left ( v_{ij} \right)$ a matrix whose rows are the left eigenvectors of $P$. Then $P$ can be written as
\[ 
P = U D V, 
\]
where $D= \left ( d_{ij} \right)$ is a diagonal matrix whose entries are the eigenvalues of the transition matrix $P$, $D= \lambda I_S$. Recall that the eigenvectors are orthogonal so that $U\,V = I_S$. Moreover, the first column of $V$ has all entries equal to 1, and if $P$ admits a steady state, the first row of $U$ is the stationary distribution. If $P$ is diagonalizable, it can be proved that the $k$--th power of $P$ can be written as 
\[
P^k = \sum_i \lambda_i^k \xi_i \, \nu_i.
\]
Since $\lambda_1 = 1$, and $|\lambda_i| < 1$, for $i=2,3,\ldots, S$,
\[
\lim_{k \rightarrow \infty} P^k = \xi_1 \nu_1,
\]
where it can be proved that the speed of convergence is given by the magnitude of $\lambda_2$, and $P$ converges faster to the steady state $\pi^*$ for smaller values of $|\lambda_2|$.  

Essentially, we model $P$ to deliver the limiting distribution $\hat{\pi}_i^*$. This is an inverse problem 
and, in general, there are infinitely many Markov matrices $P$ that give a steady state distribution $\hat{\pi}_i^*$. To single out a distribution, we use the maximum entropy principle, which postulates that given partial information
about a random variable we should choose that probability distribution for it, which is consistent with the given information,
but has otherwise maximum uncertainty associated with it \citep{Kapur:1989}. The resulting estimates are the least biased or maximally uncommitted with respect to missing information.  The maximum entropy principle is derived from information theory, originating in the work of \cite{Shannon:1948} and has been justified in numerous applications, such as matrix estimation including the estimation of transition probability matrices \citep{SchZen:1990}. For applications to image reconstruction, economics, and other areas see \cite[ch. 9]{CenZen97}. We therefore estimate the Markov matrix $P$ that satisfies the above properties while maximizing Shannon's entropy, by solving:  
\begin{align}
\mathop{\mbox{\textrm{Maximize}}}\limits_{p_{ij}}  \quad & -\sum_{ij} p_{ij} \log p_{ij} \label{eq:MaxEntropy}\\
\text{s.t.}\quad\quad\quad&\notag\\
\sum_k u_{ik} v_{kj} &= \delta_{ij}, &\mbox{for all } i,j \in \mathcal{R}, \label{eq:OrthoNormalCon}\\
\sum_k u_{ik} \, d_{kk} \, v_{kj} &= p_{ij}, &\mbox{for all } i,j \in \mathcal{R}, \label{eq:MarkovMatrixDefinition}\\
\sum_j p_{ij} &= 1, &\mbox{for all } i \in \mathcal{R}, \label{eq:StochasticMatrixCon}\\
p_{ij} &\geq 0, &\mbox{for all } i,j \in \mathcal{R}, \label{eq:NonnegativityCon}
\end{align}
where $u_{i1} = 1$, for all $i \in \mathcal{R}$, is the constraint defining the right eigenvector associated with $\lambda_1$. Constraints $v_{1j} = \hat{\pi}_j^*$, for all $j \in \mathcal{R}$, ensure that the left eigenvector associated with $\lambda_1$ is equal to the empirically estimated steady-state distribution. Eqn.~(\ref{eq:MaxEntropy}) is obtained from the additivity property of Shannon's entropy, i.e., the conditional entropy $H(X_k|X_{k-1})$ is  calculated as
\begin{equation}\label{CONDentropy}
H(X_k|X_{k-1})=\sum_{i}H(X_k|X_{k-1}=i)=\sum_{i}\left[-\sum_{j}p_{ij}\log p_{ij}\right]=-\sum_{ij} p_{ij} \log p_{ij}.
\end{equation}

This is a small scale quadratically constrained nonlinear optimization problem. The number of variables is equal to the number of regimes squared, i.e., 25 for Germany and Italy, and 9 for Greece as identified by the Bai-Perron tests (Table~\ref{tab:meanCDS}). It can be solved using off the shelf packages, such as CONOPT \citep{Drud:2005} used in our numerical results.

The eigenvalues of $P$ are set to some arbitrary values, recalling that $d_{11} = \lambda_1= 1$ and $d_{11} > d_{22} > \ldots > d_{SS}$. The possibility to arbitrarily set the eigenvalues of $P$ allows control on the expected number of time steps that the process spends consecutively in the same state. The trace of a matrix is invariant under rotation, which implies 
\begin{equation}\label{egandpii}
S \ge \sum_{i=1}^S\lambda_i=\sum_{i=1}^S p_{ii}.
\end{equation}
The expected number of consecutive time steps, $E(D_i)$, that the process spends on state $i$ is
\begin{equation}\label{EXPECduration}
E(D_i)=\sum_{k=1}^{\infty} k p_{ii}^{k-1} (1-p_{ii})=\frac{1}{1-p_{ii}}.
\end{equation}
Eqn.~(\ref{egandpii}) indicates that the average of eigenvalues is equal to the average value of $p_{ii}$ over the allowed states. Therefore, if one sets eigenvalues $\lambda_2,\lambda_3,...,\lambda_S$ close to 1, then also the average value of $p_{ii}$ turns out to be close to 1, and, according to eqn.~(\ref{EXPECduration}), the expected number of time steps that the process consecutively spends on a given state is large on average. On the contrary, if eigenvalues $\lambda_2,\lambda_3,...,\lambda_S$ are small, then probabilities $p_{ii}$ are also small.      

\begin{figure}
\includegraphics[width=\textwidth,height=0.6\textwidth,clip]{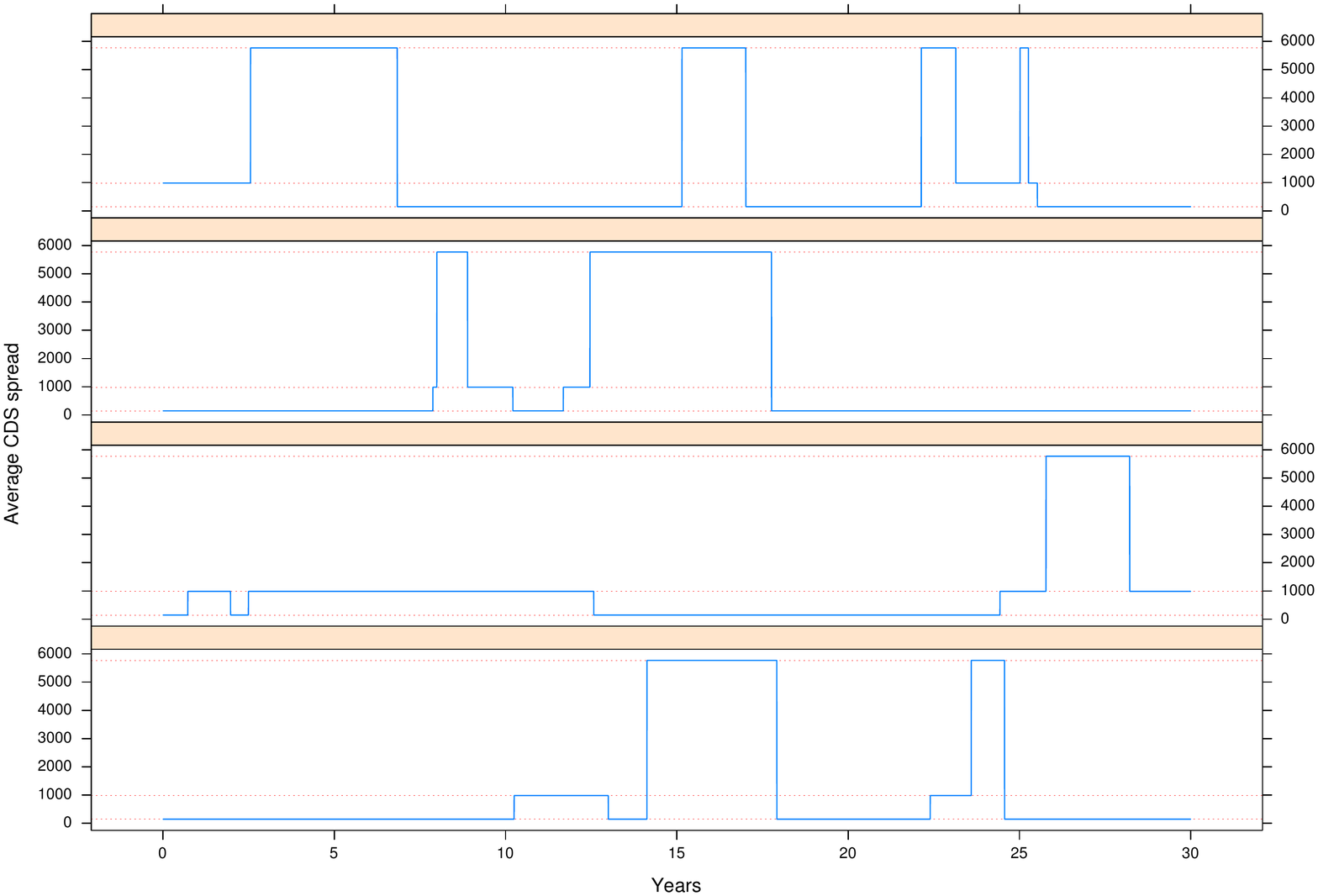}
\caption{Simulation of regimes on a daily basis over a 30-yr horizon for Greece. Solid lines illustrate the Markov process switching regimes, and a regime is defined by the average spread (dotted lines) estimated from historical data for each regime.}\label{fig:Greece-Regimes}
\end{figure}

Figure~\ref{fig:Greece-Regimes} displays four regime scenarios for Greece, generated by simulating a Markov chain with daily frequency over a 30-yr horizon. We generate scenarios based on the means spread value of the three regimes from Table~\ref{tab:meanCDS}, set monotonically decreasing eigenvalues close to 1 to obtain reasonable persistence in each regime, and obtain the empirical steady-state distribution using  (\ref{eq:EmpiricalSteadyState}). Solving (\ref{eq:MaxEntropy})--(\ref{eq:NonnegativityCon}) we obtain the following transition matrix for Greece's regimes:
\[
P =\left( \begin{array}{ccc}
0.9982	&	9.62\e{-4}	&	7.89\e{-4}\\
8.03\e{-4}	&	0.9985	&	6.56\e{-4}\\
2.62\e{-4}	&	2.76\e{-4}	&	0.9995
\end{array}\right).
\]

\begin{figure}
\includegraphics[width=\textwidth,height=0.7\textwidth,clip]{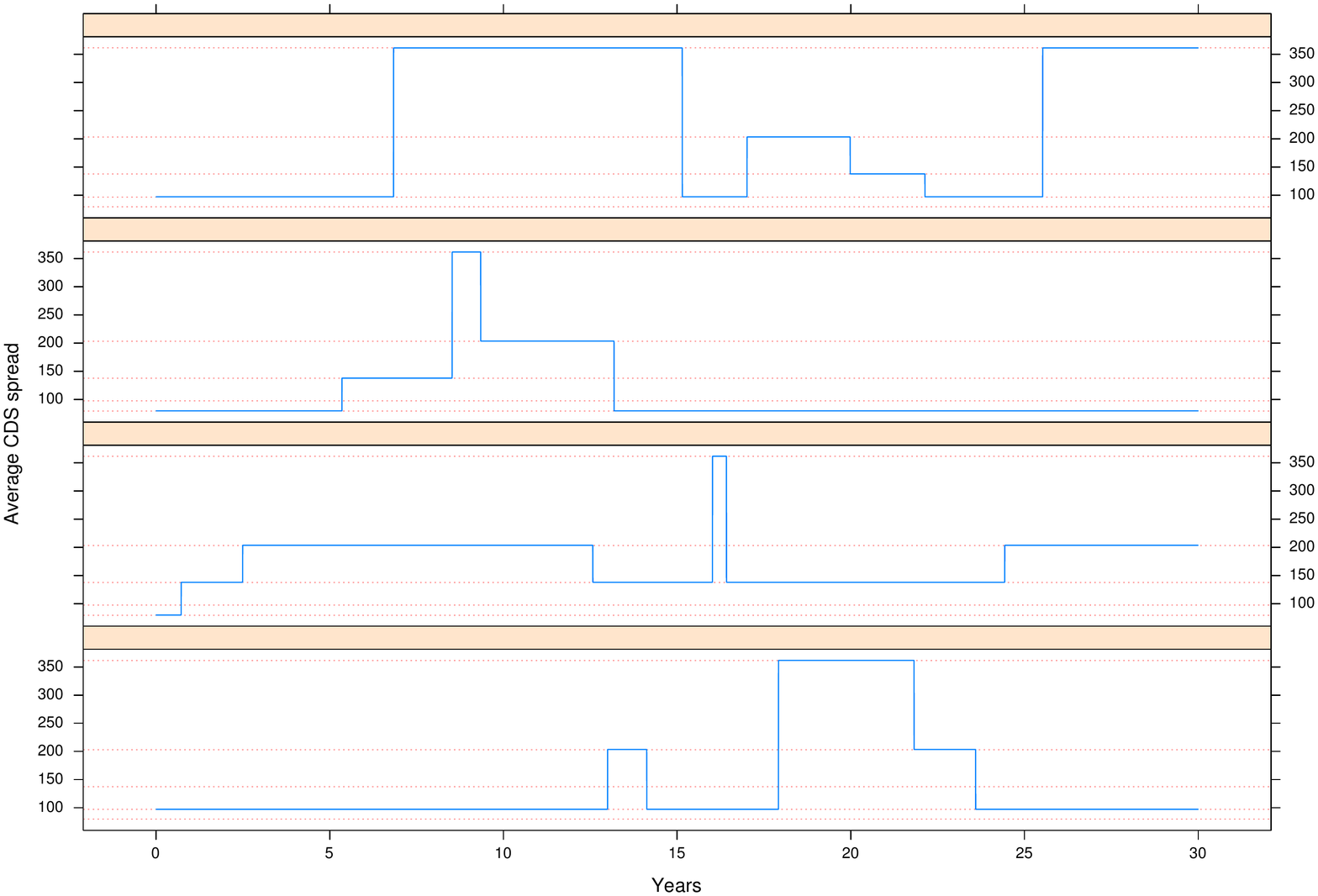}
\includegraphics[width=\textwidth,height=0.7\textwidth,clip]{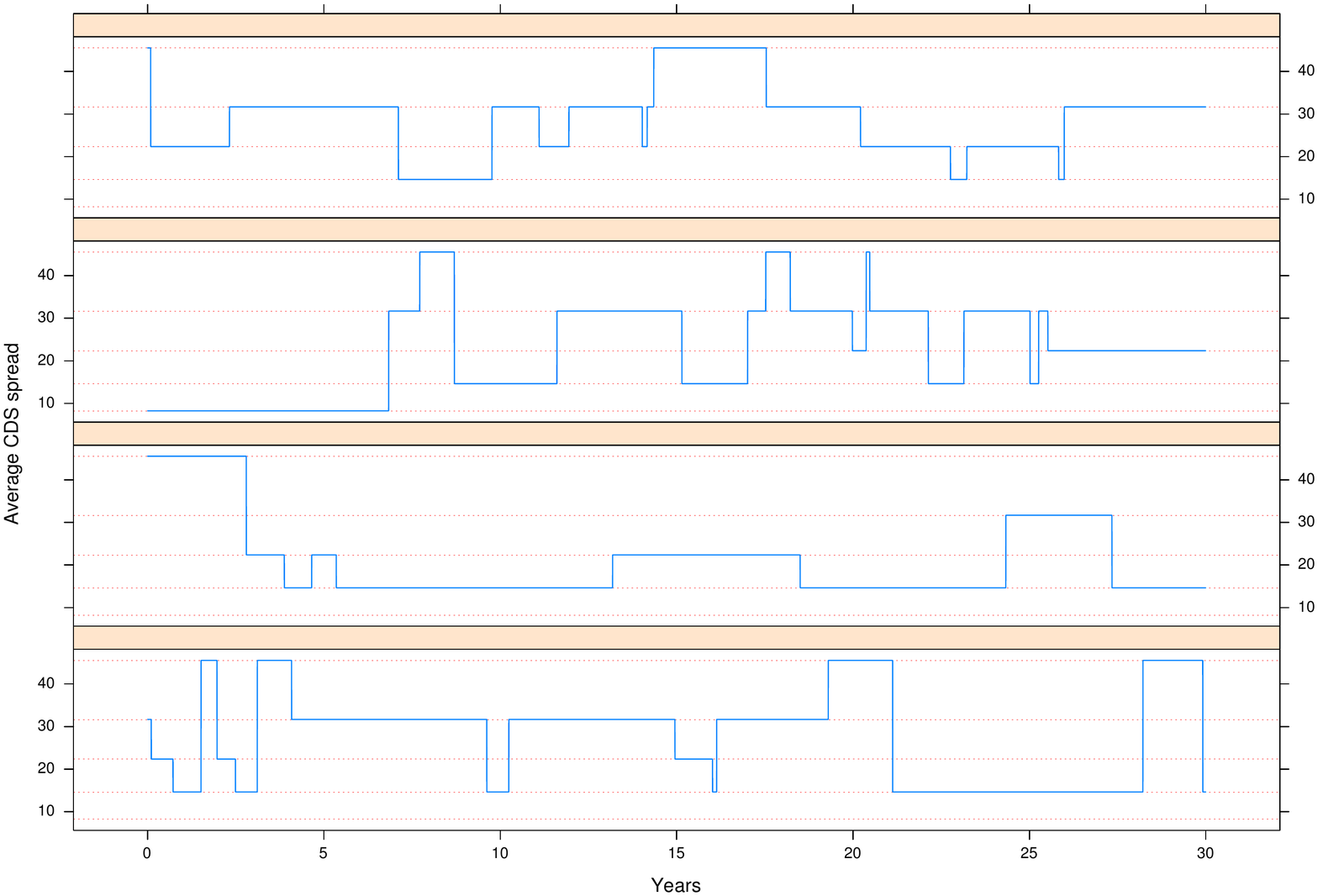}
\caption{Simulation of regimes on a daily basis over a 30-yr horizon for Italy (top) and Germany (bottom).  Solid lines illustrate the Markov process switching regimes, and a regime is defined by the average spread (dotted lines) estimated from historical data for each regime.}\label{fig:IT-DE-Regimes}
\end{figure}

We also calibrate the model for a country with less volatile spreads (Italy) and for a stable environment (Germany) and  observe similar results (see Figure~\ref{fig:IT-DE-Regimes}). Note that for Germany the mean levels of the empirically observed regimes are close to each other and modeling regime switching is not necessary. (Of course, a user may specify extreme scenarios for a German spread crisis.)

\subsection{CDS and interest rate process}

We now superimpose the CDS spread process on the spread mean regimes generated by the Markov process. Broadly speaking, we generate scenarios of CDS spreads around the regime dynamics. We need a mean-reverting process that reverts to the mean CDS spread of the (simulated) regime. Furthermore, the variance should be bounded and the spread should be non-negative.   
The SRMR model of \cite{Donoghue:2014} for CDS spread returns belongs to the class of Ornstein-Uhlenbeck processes and has the nice property that the variance of the log-returns is bounded with time, thus providing a process that does not deviate excessively from its expected value for long intervals and remains non-negative. In Appendix \ref{app:asymptotic}, we derive the conditions on the parameters of this model so that asymptotically it converges to the regime mean values. Thus, we calibrate a stochastic process that has the desirable empirically observed properties of CDS spreads and spread returns, and conforms to the regime switches.  With this approach the process dynamics capture not only the long-term mean spread but also spread and spread return volatility in each regime. Furthermore, as explained in the Appendix, this process allows to calibrate short term fluctuations and hence the smoothness of the curve.

\begin{figure}
\includegraphics[width=\textwidth,height=0.6\textwidth,clip]{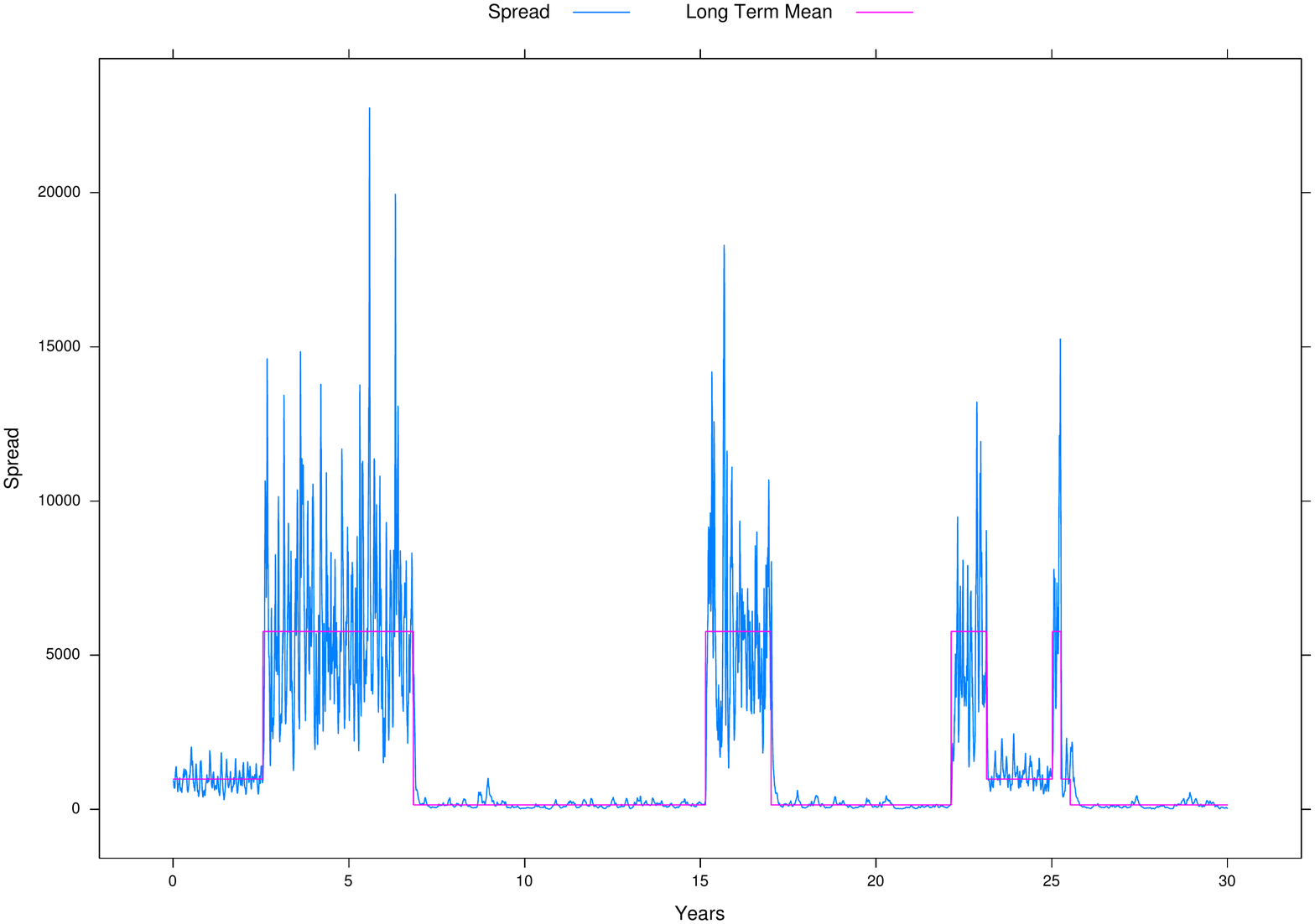}
\includegraphics[width=\textwidth,height=0.6\textwidth,clip]{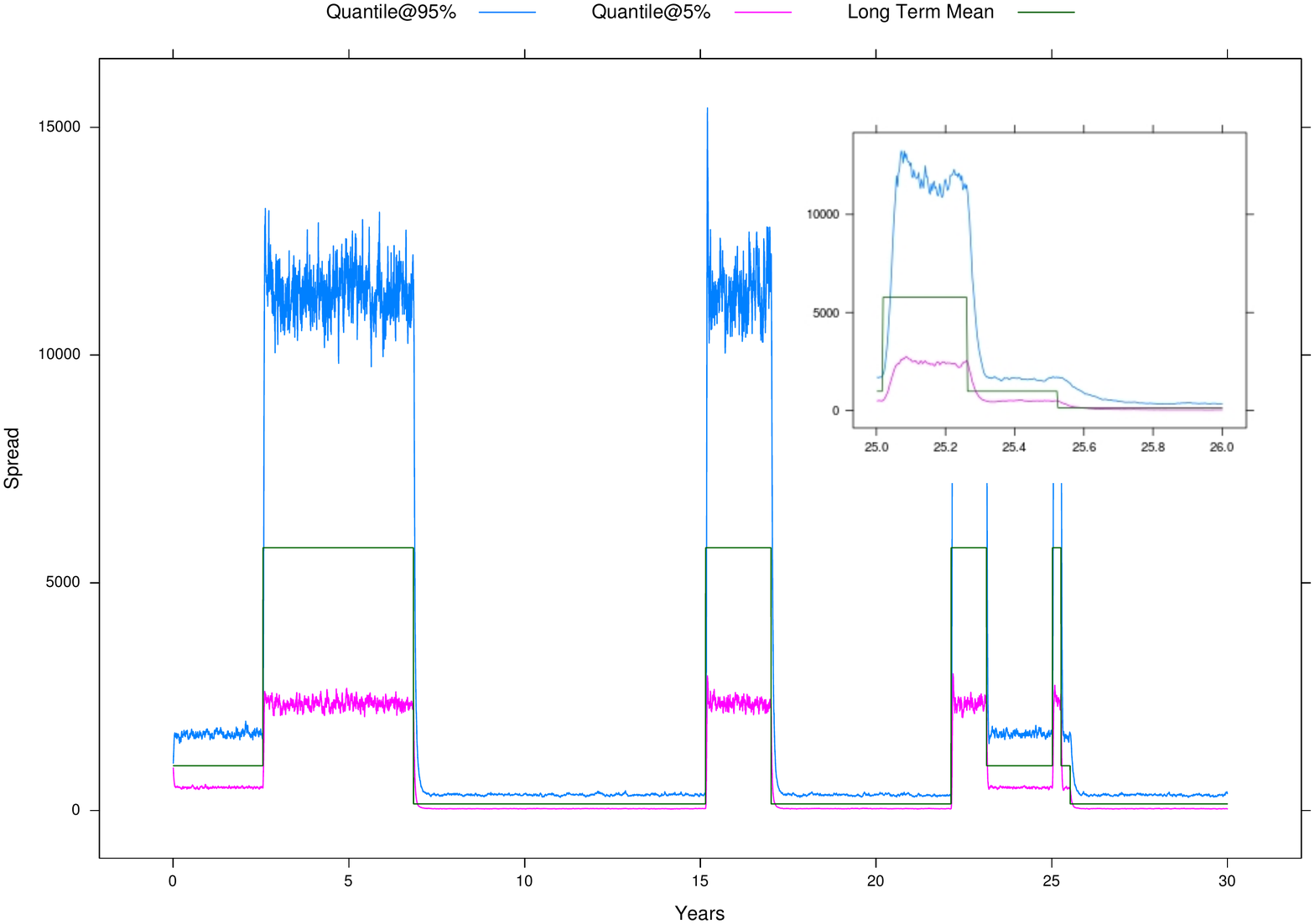}
\caption{A typical simulation path of the daily CDS spread for Greece over a 30-yr horizon (top) and the 5\% and 95\% quantiles over 1000 scenarios (bottom) with a detail of the spread dynamics between years 25 and 26. The average spread of each regime is as estimated in Table~\ref{tab:meanCDS}.}\label{fig:GreeceSpreadSimulation-30}
\end{figure}

\begin{figure}
\includegraphics[width=\textwidth,height=0.6\textwidth,clip]{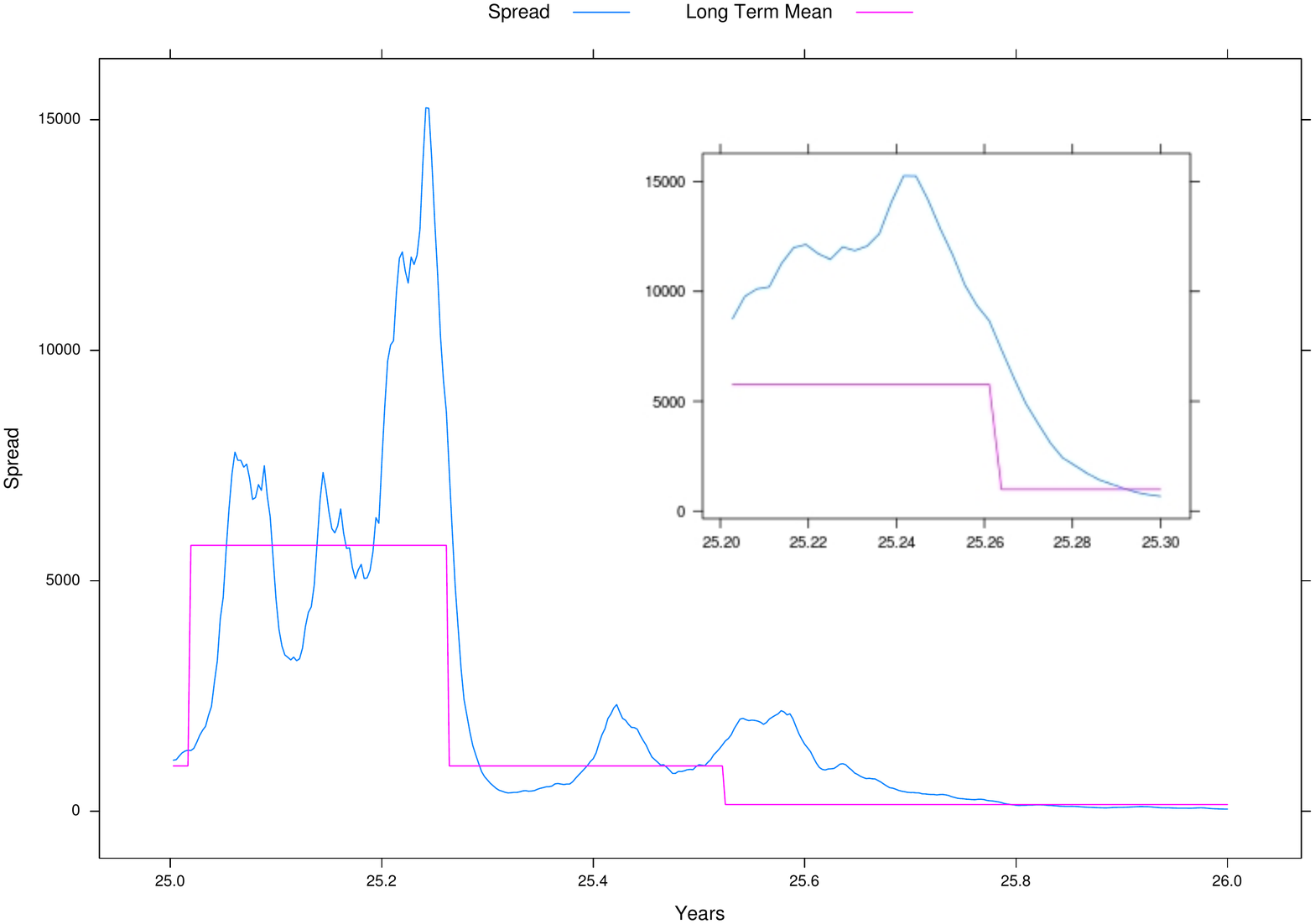}
\caption{Simulated daily CDS spread for Greece during regime switching from year 25 to 26.
}\label{fig:GreeceSpreadSimulation-30-Zoom-25-26}
\end{figure}

Figure~\ref{fig:GreeceSpreadSimulation-30} illustrates a sample scenario of Greek CDS spread, around the regime scenario from Figure~\ref{fig:Greece-Regimes} (top panel). The simulation is run on a daily basis over a 30-yr horizon. The process follows the mean CDS spread level for each regime. A first impression is of a process with unrealistic jumps of the spread coinciding with regime switching. Moreover, the dynamics of the spread for the tranquil regime appear to be flat, with negligible volatility. This is due to y-axis scaling to capture the wide range of spreads for Greece over a long horizon. Zooming in at the simulated series we observe a smooth transition between regimes, with higher volatility even in the tranquil regime. Figure~\ref{fig:GreeceSpreadSimulation-30-Zoom-25-26} displays the spread dynamics between years 25 and 26, where there are three consecutive regime transitions. Transition from crisis to turbulent regime is abrupt, but the spread changes with a reasonable gradient, as seen in the inset of the figure.

One desirable property of the model is the bounded variance of the stochastic process. Figure~\ref{fig:GreeceSpreadSimulation-30} (bottom) illustrates the 5\% and 95\% quantiles of the CDS spreads obtained over 1000 simulations. We observe that volatility does not increase with time and is dependent only on the given regime, so that turbulent regimes have higher volatilities than tranquil regimes and crisis regimes even higher. The largest Greek CDS spread during the crisis was almost 15,000, and was generated by our simulation at the 95\% quantile.    

As explained in the next session, the S-CoCo cashflows are discounted using the EURO AAA-rated bond yields (E-AAA for short). We simulate the E-AAA short rate dynamics following the approach just described. To this purpose, we extract from the historical series of the E-AAA yield curve the series of the 1-month rate, and we determine the regime sub-intervals and relative statistics to calibrate the model. We remark that this implementation does not match the  term structure, and, therefore, we are not able to match observed bond prices on a given date. A workaround to this drawback would be to use a time-dependent process matching the actual forward curve, and calibrating the parameters of the model with given volatilities (implicit or historical ones). That is, unlike our implementation, where the process fluctuates around the simulated regimes, we could make the short rate to mean-revert towards an exogenously given forward curve.

\section{Modeling sovereign contingent convertible debts} \label{s:modeling}

We develop now the pricing models using Monte Carlo simulations.  Prices are obtained as the expected discounted cashflows from simulations of the Markov chain and the stochastic process of spreads and interest rates in each regime. We also show how to obtain state contingent prices at some risk horizon to facilitate risk management.

\subsection{Pricing}\label{sec:CoCoPricing}

We denote by $\xi=\left\{r_t, s_t\right\}$ the coupled stochastic process of the short rate $r_t$ and CDS spread $s_t$, where we assume that $\mbox{\textrm{cov}}\left[r_t,s_t\right]=0$.\footnote{We can also have correlated processes $r_t$ and $s_t$. In this case, the covariance matrix will be factorized through a Cholesky decomposition, and standard Montecarlo sampling for correlated processes will be used to simulate jointly spread and interest rate, see the Appendix.}

To simplify  notation, we use $t$ to indicate discrete time steps, from the index set $\mathcal{T}=\{0,1,2, \ldots, T\}$. We draw from the probability distribution of $\xi$ a discrete number of sample paths (\textit{scenarios}), $\xi^l = \left\{r^l_{t}, s^l_{t}\right\}$, where $l \in \Omega=\{1,2, \ldots, N\}$ and $t \in \mathcal{T}$. The time-discretized approximation of the stochastic process $\xi$, for each scenario $l \in \Omega$, is obtained from eqn.~(\ref{eq:StochasticSpreadReturnDynamics}) by drawing $N$ random samples of the diffusion term $w_{t}$ from a Gaussian distribution. All sampled scenarios are equally likely with probability $1/N$, and large number of scenarios approximate the underlying Gaussian distribution. This is a standard Monte Carlo simulation, and converged well in our numerical implementation. For advanced variance reduction techniques see \citep[ch. 4]{Glasserman03}.

Denote by $\bar{s}$ the threshold of the CDS spread which activates the standstill. If at time $t$ and under scenario $l$ the CDS rate $s^l_{t}$ hits $\bar{s}$, coupon payments are suspended for the next $K$ periods. We define the set of time periods with payment standstill by $\mathcal{T}^l_m = \left \{ t, t+1, \ldots, t+K \right\}$, and $m=1,2,\ldots,M$, where $M$ is the number of times that the standstill mechanism is activated under scenario $l$, with the following properties:
\begin{enumerate}
	\item For any $m$ and $n$, $\mathcal{T}^l_m \cap \mathcal{T}^l_n = \emptyset$, to preclude overlapping of payment standstills.
    \item For any $\tau \in \mathcal{T}^l_m$, $\tau > t$, if $s^l_{\tau} \geq \bar{s}$ the trigger signal is ignored, to avoid multiple triggering during a standstill interval.
\end{enumerate}

The set of periods $t \in \mathcal{T}$ with payment standstill for scenario $l$ is
\begin{equation}\label{eq:triggeringSet}
\Lambda^l = \bigcup_{m=1}^M \mathcal{T}^l_m,
\end{equation}
and we define an indicator function $\mathbbm{1}_{\Lambda^l}: \mathcal{T} \rightarrow \lbrace 0,1 \rbrace$ as
\begin{equation}
\mathbbm{1}_{\Lambda^l}(t) = 
\left\lbrace
\begin{array}{ll}
0,&\mbox{if}\quad t \in \Lambda^l\\
1,&\mbox{if}\quad t \notin \Lambda^l.
\end{array}
\right. 
\end{equation}
The standstill provision includes a special treatment of credit events occurring within $K$
periods before maturity. In such cases, coupon payment standstill implies deferral of principal payment. In particular, denoting by $\mathcal{T}^l_Z$ the terminal standstill set under scenario $l$, and defining by $J^l$ the first time step of  $\mathcal{T}^l_Z$, $J^l=\min{\mathcal{T}^l_Z}$, the principal payment is delayed by $\Delta T^l=T - J^l + 1$, provided that $T-J^l < K$.

The S-CoCo price is obtained as the expectation, over scenarios $l \in \Omega$, of the present value of coupon and principal payments. That is, 

\begin{equation} \label{eq:cocoPricing}
P_0 = \frac{1}{N}\sum_{l \in \Omega} \sum_{t \in \mathcal{T}} B^l(0,t) \mathbbm{1}_{\Lambda^l}(t) \; c + B^l(0,T+\Delta T^l),
\end{equation}
where  $c$ is the coupon and $B^l(t,s)$ is the discount factor between time periods $t$ and $t'$
\begin{equation}
B^l(t,t') = \exp\left\lbrace - \sum_{u=t}^{t'} r^l_{u}\right\rbrace.
\end{equation}

There can be several variants of the standstill provision, such as payment standstill with an associated maturity extension for as long the spread exceeds the threshold. Also, there are various alternative ways to treat coupon payments missed during the standstill, such as resumption of nominal value payments until the (extended) maturity ---this was our original S-CoCo suggestion--- or resumption of payments on an accrual basis or total write-down of missed payments. The pricing formula still applies but modifications are needed of the definition of the triggering set $\Lambda^l$ or a different accounting of cashflows in (\ref{eq:cocoPricing}). Modifications are conceptually straightforward but complicate the notation and we do not give them here. 

Using numerical line search we solve pricing formula (\ref{eq:cocoPricing}) for $c$ such that
\begin{equation}\label{eq:ParRate}
P_0(c) = 1.
\end{equation}
The difference between $c$ and the par rate of a AAA sovereign bond is the premium charged by investors to buy the  S-CoCo.

\begin{figure}
\includegraphics[width=\textwidth,height=0.6\textwidth,clip]{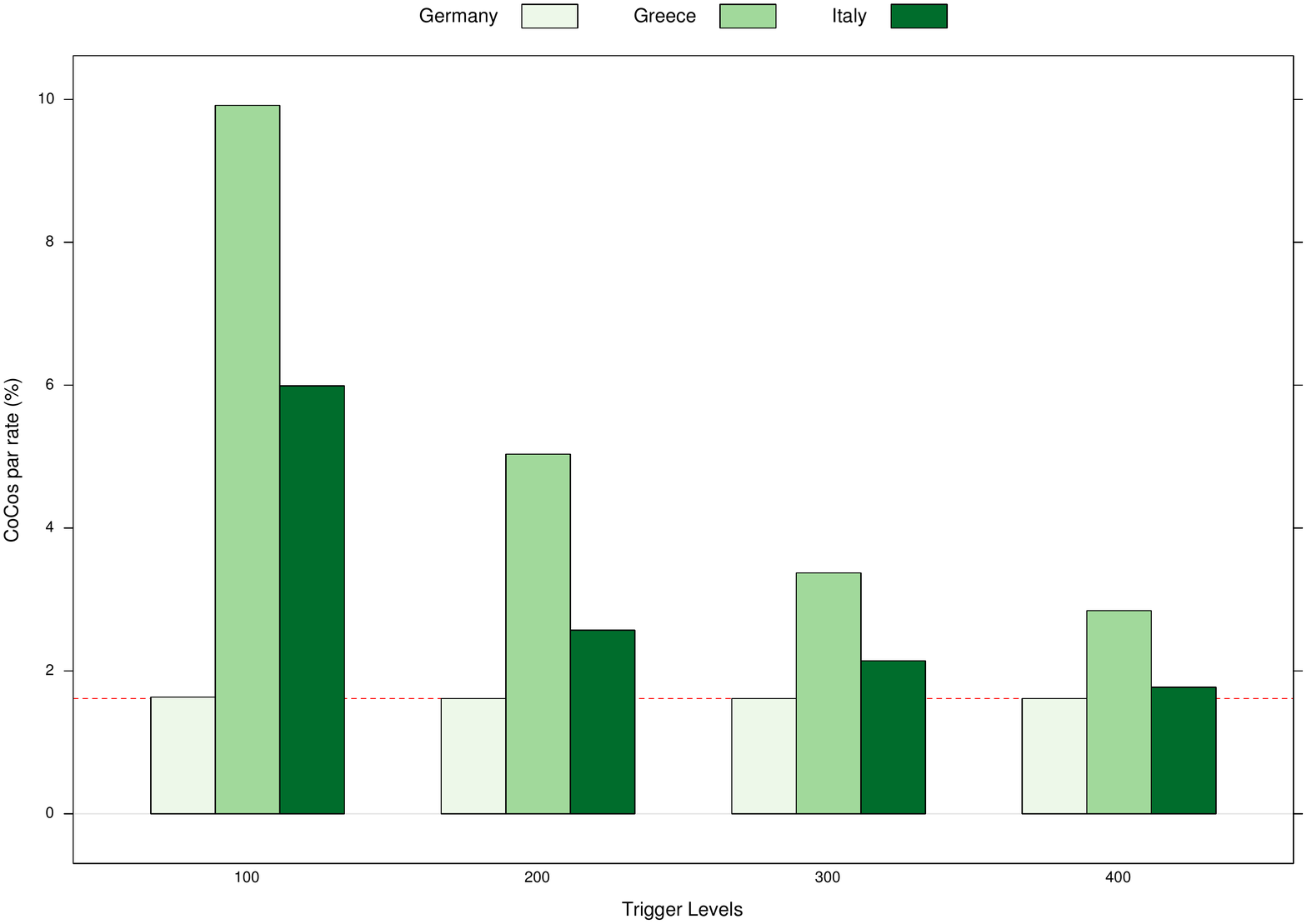}
\caption{Par rate of the S-CoCo {\em vs} trigger thresholds $\bar{s}$. The red dashed line indicates the par yield of a AAA-rated bond (1.6\%).}\label{fig:CoCoParRates}
\end{figure}

Figure~\ref{fig:CoCoParRates} displays par rates of a 20-yr S-CoCo for the three countries of our study with threshold $\bar{s}=100, 200, 300, 400$. The CDS processes are calibrated on daily historical series from January 2007 to the end of 2016, except for Greece whose CDS trading was suspended at the end of 2012. The parameters of the short rate dynamics are inferred from the daily historical series of the E-AAA 1-month bond yield.
Each par rate is computed by solving eqn.~(\ref{eq:ParRate}) over a set of 100 regime scenarios and 1000 interest rates and spread scenarios for each regime, for a total of 100,000 paths of length 20 years and semi-annual time step. Also shown in the figure is the par rate of a plain AAA-rated bond (1.6\%). Greece has the highest premium over the AAA-rated yield due to the very high average level of CDS spreads of the recent past. The premium increases as $\bar{s}$ is reduced since the probability of breaching the threshold increases with a commensurate increase in the number of standstill time periods. German S-CoCo is priced at par with AAA-rated bonds as the likelihood of German CDS spreads breaching even a very low threshold is virtually nil. The Italian spread is, naturally, between Greece and Germany. 

Note that, the convergence of the par rate to the AAA level is due to the unique short rate dynamics used for all countries, which is calibrated on the AAA-rated bond historical yield series. Differentiation of the minimum par rate, and therefore convergence to different minimum levels as $\bar{s}$ increases, would be observed if the short rate dynamics are calibrated separaibliograptely for each country.

We illustrate the sensitivity of price and par rate estimates to changes of the regime probabilities. The experiment is for the Greek case where the Bai-Perron test identifies three regimes with estimated steady-state probability vector $\hat{\pi}^* = \left( 0.5612,0.2888,0.15\right )$. We perturb $\hat{\pi}_i^*$ by sampling a Dirichlet distribution with parameters $\alpha \hat{\pi}^*$ \citep{KBJ:2005}, where $\alpha$ is the \textit{concentration parameter} determining how concentrated is the probability mass of a Dirichlet distribution around the given discrete probability distribution $\hat{\pi}^*$.  The support of the Dirichlet distribution $\mathbbm{D}(\alpha \pi)$ with $\pi \in [0,1]^N$, is the set of $N$-dimensional vectors $p=\left ( p_1 \, p_2 \, \ldots \, p_N \right )$ whose entries are real numbers in the interval (0,1), and the sum of the entries is 1. Equivalently, the domain of the Dirichlet distribution is itself a set of probability distributions, namely the set of $N$-dimensional discrete distributions.

\begin{figure}
\includegraphics[width=\textwidth,keepaspectratio,clip]{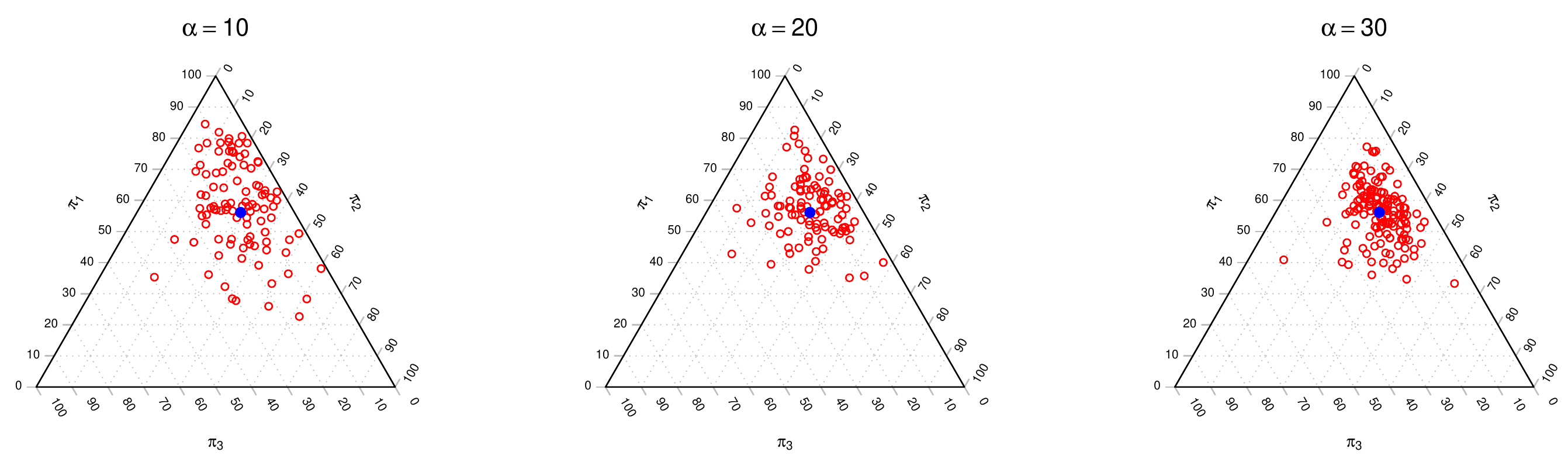}
\caption{Samples drawn (red circles) from a Dirichlet distribution $\mathbbm{D}(\alpha \hat{\pi}^*)$ for different concentration parameters $\alpha$. Higher values of $\alpha$ indicate more confidence about the estimate $\hat{\pi}^*$ (blue circle) and the samples are more concentrated around the estimate.}\label{fig:TernaryPlot}
\end{figure}

In Figure~\ref{fig:TernaryPlot}, we show the samples drawn from $\mathbbm{D}(\alpha \hat{\pi}^*)$ where the concentration parameters are set to $\alpha = 10, 20, 30$. $\alpha$ can be viewed as the confidence of the decision maker about the estimate $\hat{\pi}^*$ of the steady state discrete distribution ${\pi}^*$, with higher $\alpha$ denoting more confidence that $\hat{\pi}^*$ is indeed a good estimate of the true ${\pi}^*$.

\begin{figure}
\begin{center}
\includegraphics[width=0.45\textwidth,keepaspectratio,clip]{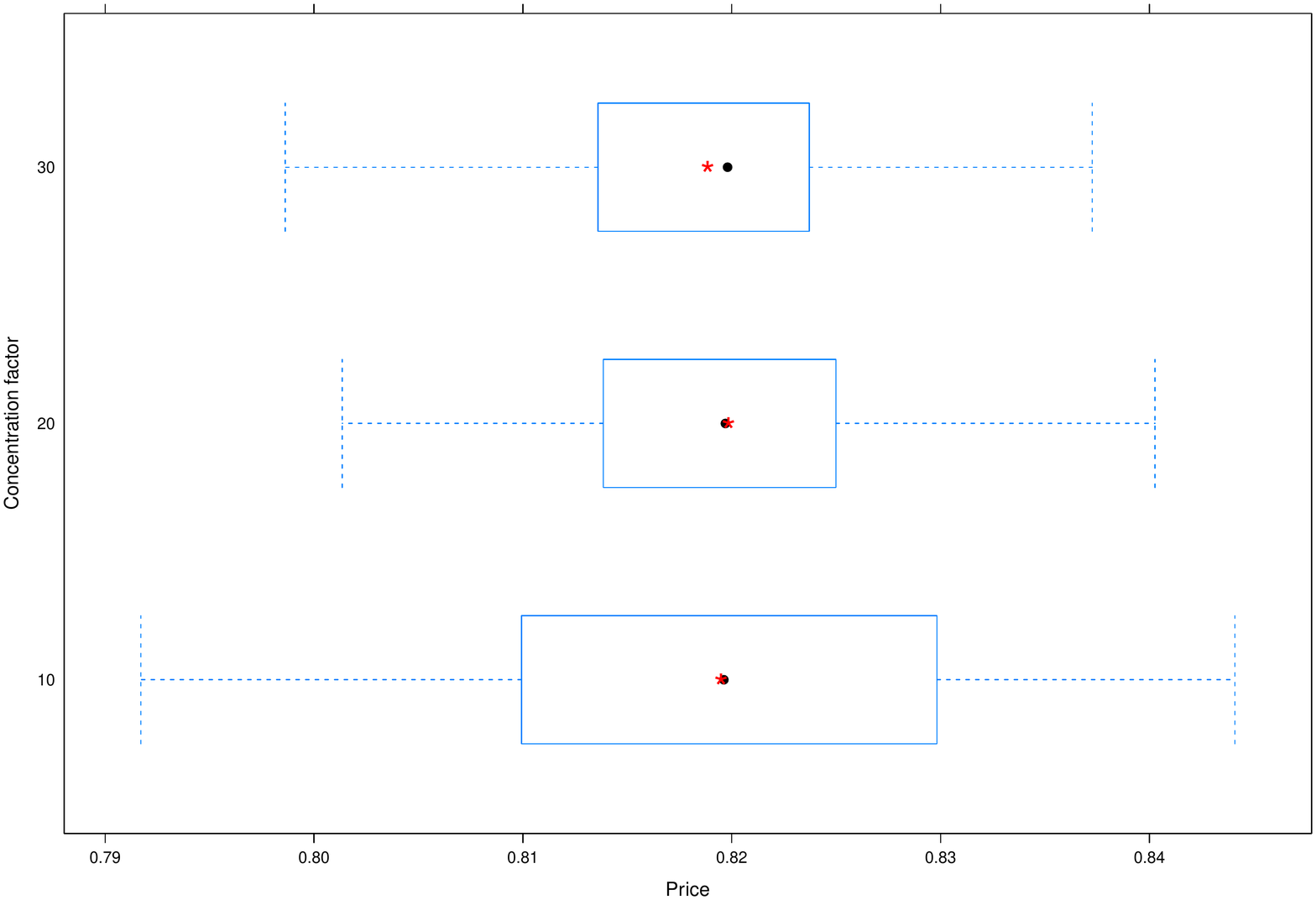}
\includegraphics[width=0.45\textwidth,keepaspectratio,clip]{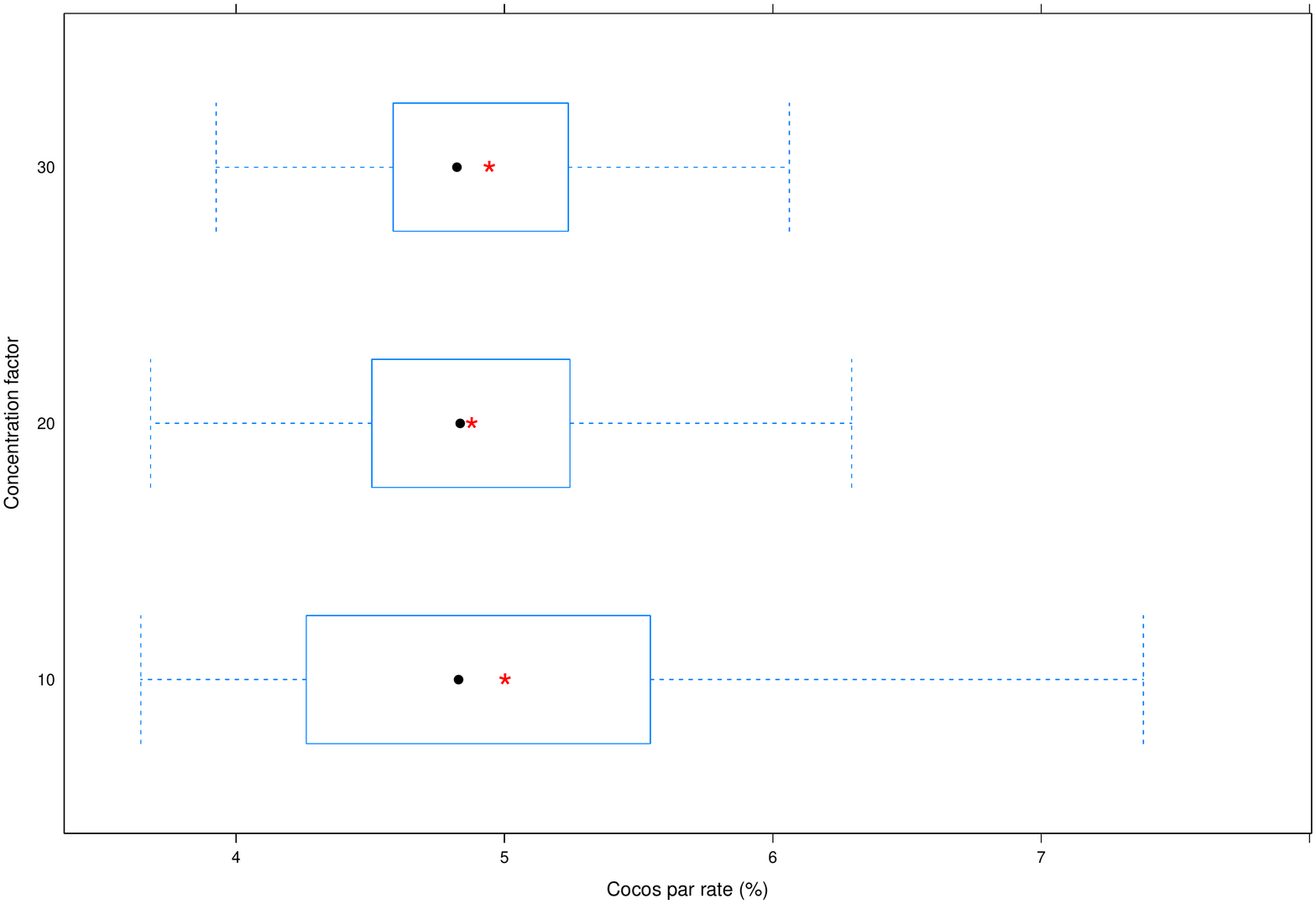}    
\end{center}
\caption{Box-whisker plots of the prices (left panel) and par rates (right panel) for a  20-year Greek S-CoCo with threshold 200. The perturbation of the estimated steady state distribution generates relatively stable prices but higher variability of par rates.}\label{fig:CoCosPriceSensitivity}
\end{figure}

Given the sampled probability distributions $p^l$, with $l=1,2,\ldots 100$, we estimate the price of a 20-year S-CoCo with threshold 200. The distribution of prices is displayed using box-whisker plots, where the box delimits the inter-quartile range from the 25\% to the 75\% quantiles, whereas the black dot and the red star are, respectively, the median and the average price. The box-whisker plot of the CoCos price and par rates are displayed in Figure~\ref{fig:CoCosPriceSensitivity}. The inter-quartile range is quite stable for price estimate, ranging from 1.1\%  for high concentration value to 2.4\% for low value. Higher sensitivity is displayed by par rates, with changes ranging from 60bp for high concentration values to 126bp for low concentration,  due to the nonlinear relation between par rates and prices.

\subsection{State contingent pricing and holding period returns}

For risk management we need the price  (equivalently, return) probability distribution of financial instruments at the risk horizon to compute risk measures or for portfolio optimization  or credit value adjustments. Such distributions are conditioned on the relevant risk factors and are needed under the true, objective, probability measure. See \cite{MulZen:1994} for generation and use of these distributions for fixed income securities and \cite{ConZen:2016jgd} for use in risk management for sovereign debt restructuring.

Given the stochastic dynamics of a risk factor, a closed form expression of the expected value of the pricing function is not always available, especially when there are more than one risk factors.
Hence, we resort to the numerical \textit{Least Square Montecarlo} ---LSM in short--- of \cite{LongSchw01}. This method was developed to price American options and can be suitably modified to compute the conditional expectation of the S-CoCo bond contingent on the short rate $r_t$ and the trigger binary function $\mathbbm{1}_{\Lambda}(t)$. 

LSM is based on backward induction whereby the expected value of the (discounted) asset payoff at $t+1$ is approximated by a function of the realizations of the random variable at $t$:  
\begin{equation}
E \left[ V_{t+1}(X_{t+1}) | X_t= x_t \right] \approx f_t(x_t,\beta_t), \quad x_t \in \re^d.
\end{equation}
In the S-CoCo context, $X_t = \left( r_t, \mathbbm{1}_{\Lambda}\right)$ is the 2-dimensional vector which takes values $x_t=\left( r^l_t, \mathbbm{1}_{\Lambda^l}\right)$ obtained by the Montecarlo pricing simulation. 

The payoff function $V_{t+1}(X_{t+1})$ has to account for the cashflow occurring at $t+1$. This is made up by the possible coupon payment, plus the expected value of the S-CoCo at $t+1$. For a given realization of the random variable $X_{t+1}$, we have
\begin{equation}\label{eq:LSM-General}
V_{t+1}(r^l_{t+1}, \mathbbm{1}_{\Lambda^l}(t+1)) = \left[ \mathbbm{P}_{t+1}\left( r^l_{t+1}, \mathbbm{1}_{\Lambda^l}(t+1) \right) + c \mathbbm{1}_{\Lambda^l}(t+1) \right] B^l(t,t+1),
\end{equation}
where, $\mathbbm{P}_{t+1}\left(r^l_{t+1}, \mathbbm{1}_{\Lambda^l}(t+1)\right)$ is the regression function approximating the expected S-CoCo price in the next period and $B^l(t,t+1)$ is the discount factor.  

Starting from $V_T(x)$ (see discussion below about the terminal payoff function), we estimate backwards the parameters $\beta_t \in \re^{M}$ and the error term $\epsilon_t \in \re$ that best fist the expected value
\begin{equation}\label{eq:RegressionGeneral}
\mathbbm{P}_t(x_t,\beta_t) + \epsilon_t = E \left[ V_{t+1}(X_{t+1}) | X_t= x_t \right],
\end{equation}
where $\mathbbm{P}_j(\cdot)$ is obtained as a linear combination of basis functions
\begin{equation}
\mathbbm{P}_t(x_t,\beta_t) = \sum_{k=1}^M \beta_{tk} \phi_k(x_t).
\end{equation}
The choice and the number of basis functions $\phi_k$ depend on the characteristics of the problem under review. Most authors suggest a trial-and-error approach, starting from simple basis functions and then increase their complexity (for example, using power function with dampening factors, Hermite or Laguerre polynomials), together with statistical selection procedures to find the optimal number of functions. Following \citep[p. 462]{Glasserman03} we set $\phi_k(r_t) = r_t^k$ and $\phi_k(\mathbbm{1}_{\Lambda}(j)) = \mathbbm{1}_{\Lambda}(t)$. For the short rate we tried different sets of basis functions $\left\lbrace r_t^k \right\rbrace_{0}^{M-1}$, where $M=3,4,5$. For the binary variable, we only considered $k=1$ since any power of  $\mathbbm{1}_{\Lambda}(t)$ will deliver the same value.

Given the sample values for $r_t^l$, $\mathbbm{1}_{\Lambda^l}(t)$ and  $V^l_{t+1}$,  starting from $j=T$ and  proceeding backwards until $t=1$, we estimate $\left\lbrace \beta_{tk} \right\rbrace_{k=0}^M$ through standard OLS. The price of the S-CoCo at $t=0$ is given by
\begin{equation}
P_0^{\mbox{\tiny LSM}} = \frac{1}{N} \sum_{l \in \Omega} \left\lbrace \left[ \mathbbm{P}_{1}(r^l_{1}, \mathbbm{1}_{\Lambda^l}(1)) + c \mathbbm{1}_{\Lambda^l}(1) \right] B(0,1) \right\rbrace.
\end{equation}

As discussed in Section~\ref{sec:CoCoPricing}, the standstill provision allows for principal payment to be postponed if the triggering event occurs within $K$ periods before maturity. Therefore, at $j=T$ the value of the S-CoCo is contingent on the scenario $l$ and is given by
\begin{equation}
V^l_T(x^l_T) = 
\left\lbrace
\begin{array}{ll}
\mathbb{B}(T,T+\Delta T^l),&\mbox{if}\quad T \in \Lambda^l\\
1+c,&\mbox{if}\quad T \notin \Lambda^l,
\end{array}
\right. 
\end{equation}
where $\mathbb{B}(T,T+\Delta T^l)$ is the expected value of a zero coupon bond maturing at $T+\Delta T^l$. To compute $\mathbb{B}(T,T+\Delta T^l)$, we apply again LSM with $r_t^l$ the only conditioning variable, for $t=T, T+1, \ldots, T+\Delta T^l$, and terminal value $V_{T+\Delta T^l}=1$. 

Table~\ref{tab:LSM-MAPE} compares the results obtained using different sets of basis functions $\left\lbrace r_t^k \right\rbrace_{0}^{M-1}$ and the dummy variable $\mathbbm{1}_{\Lambda}(j)$.
The prices $\bar{P}_0^{\mbox{\tiny LSM}}$ in Table~\ref{tab:LSM-MAPE} are average prices of an S-CoCo with 10 year maturity\footnote{A shorter maturity is used to reduce computational time, but similar results are obtained when running the experiment for a 20-yr bond on a single set of basis functions.}, obtained by changing the seed of the random engine to generate 5000 sample paths of length 10 years of short rates and CDS spreads, keeping the regimes from Section~\ref{sec:CoCoPricing}. In the same table we show the mean absolute percentage error with respect to the Montecarlo price. (For a true comparison we need a price obtained through a completely different approach, which, at the moment, is not available for S-CoCo.) The experiment highlights that basis functions up to degree two deliver satisfactory approximations. 

\begin{table}[htbp]
  \centering
    \begin{tabular}{lll}
    \toprule
          Basis functions & $\bar{P}_0^{\mbox{\tiny LSM}}$     & MAPE \\
    \midrule
    $1$, $r$, $\mathbbm{1}_{\Lambda}$     & 0.95558736 & 0.09163\% \\
    $1$, $r$, $r^2$, $\mathbbm{1}_{\Lambda}$      & 0.96610208 & 0.08913\% \\
    $1$, $r$, $r^2$, $r^3$, $\mathbbm{1}_{\Lambda}$     & 0.95644892 & 0.33191\% \\
    $1$, $r$, $r^2$, $r^3$, $r^4$, $\mathbbm{1}_{\Lambda}$      & 0.97058076 & 1.67800\% \\
    \bottomrule
    \end{tabular}%
  \caption{Average LSM price at the root node and mean absolute percentage error (MAPE) with respect to Montecarlo pricing for different basis function sets.}\label{tab:LSM-MAPE}%
\end{table}%

\begin{figure} 
\subfloat[Greece]{\includegraphics[width=\textwidth,height=0.40\textwidth,clip]{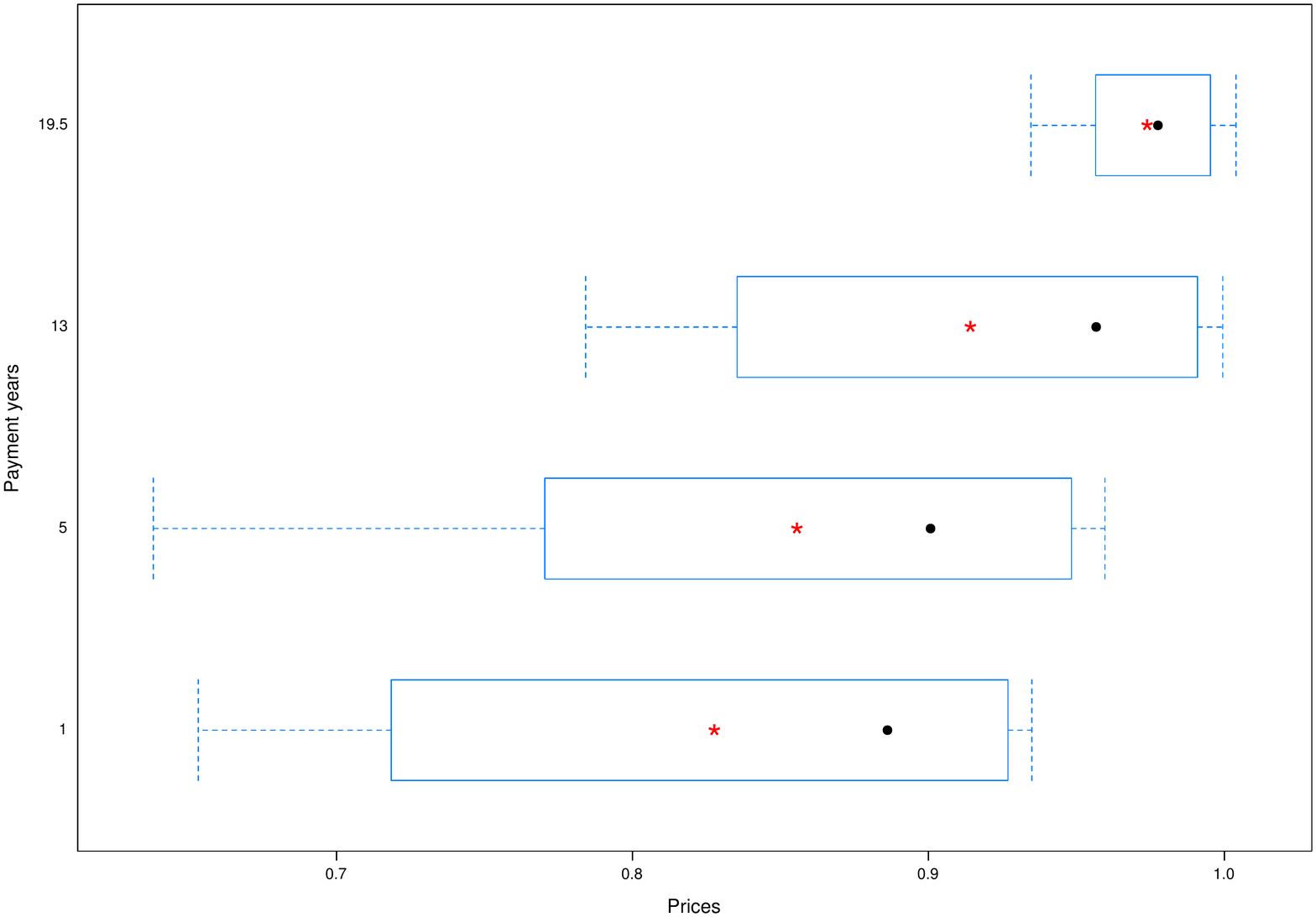}}\\
\subfloat[Italy.]{\includegraphics[width=\textwidth,height=0.40\textwidth,clip]{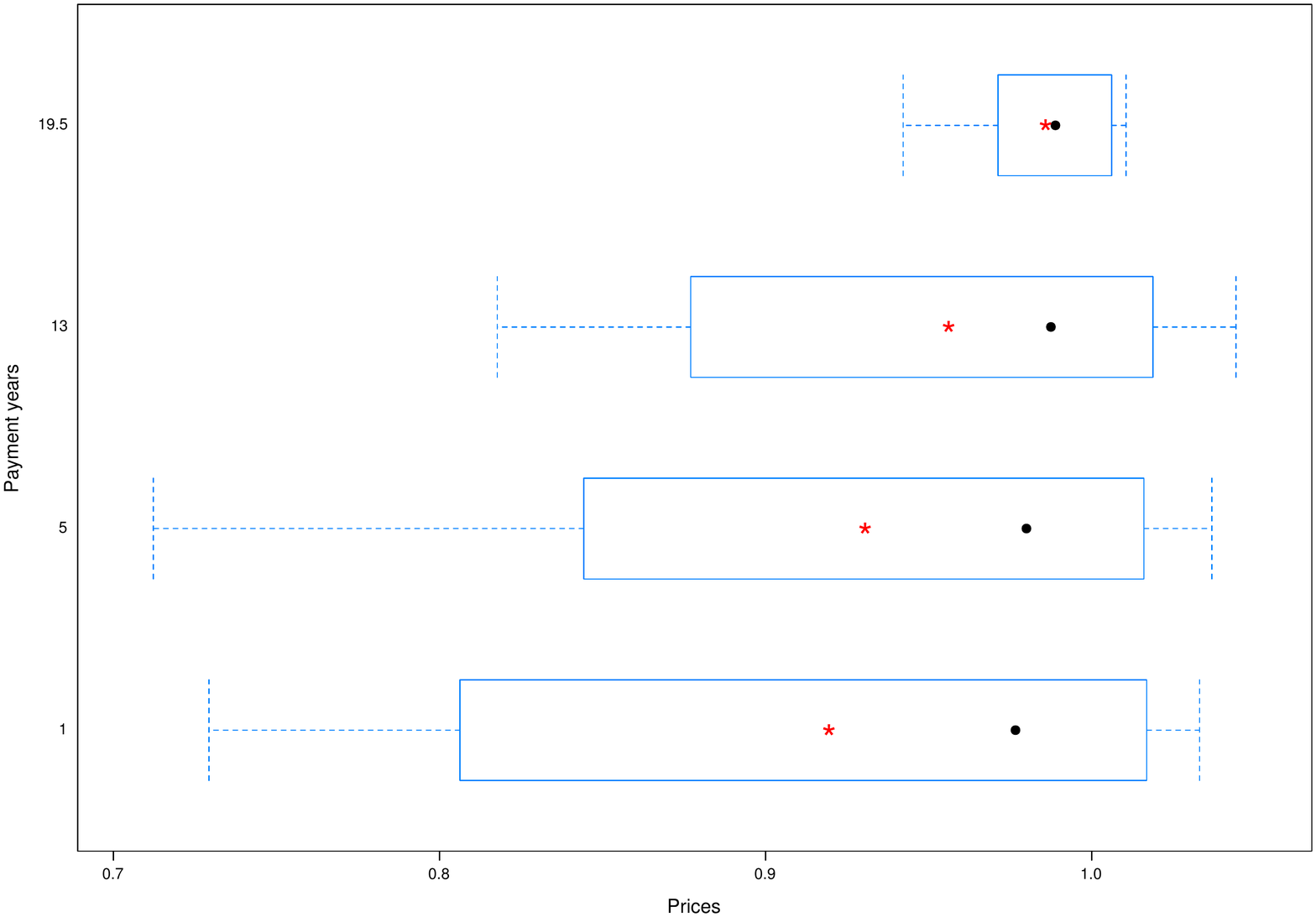}}\\
\subfloat[Germany]{\includegraphics[width=\textwidth,height=0.40\textwidth,clip]{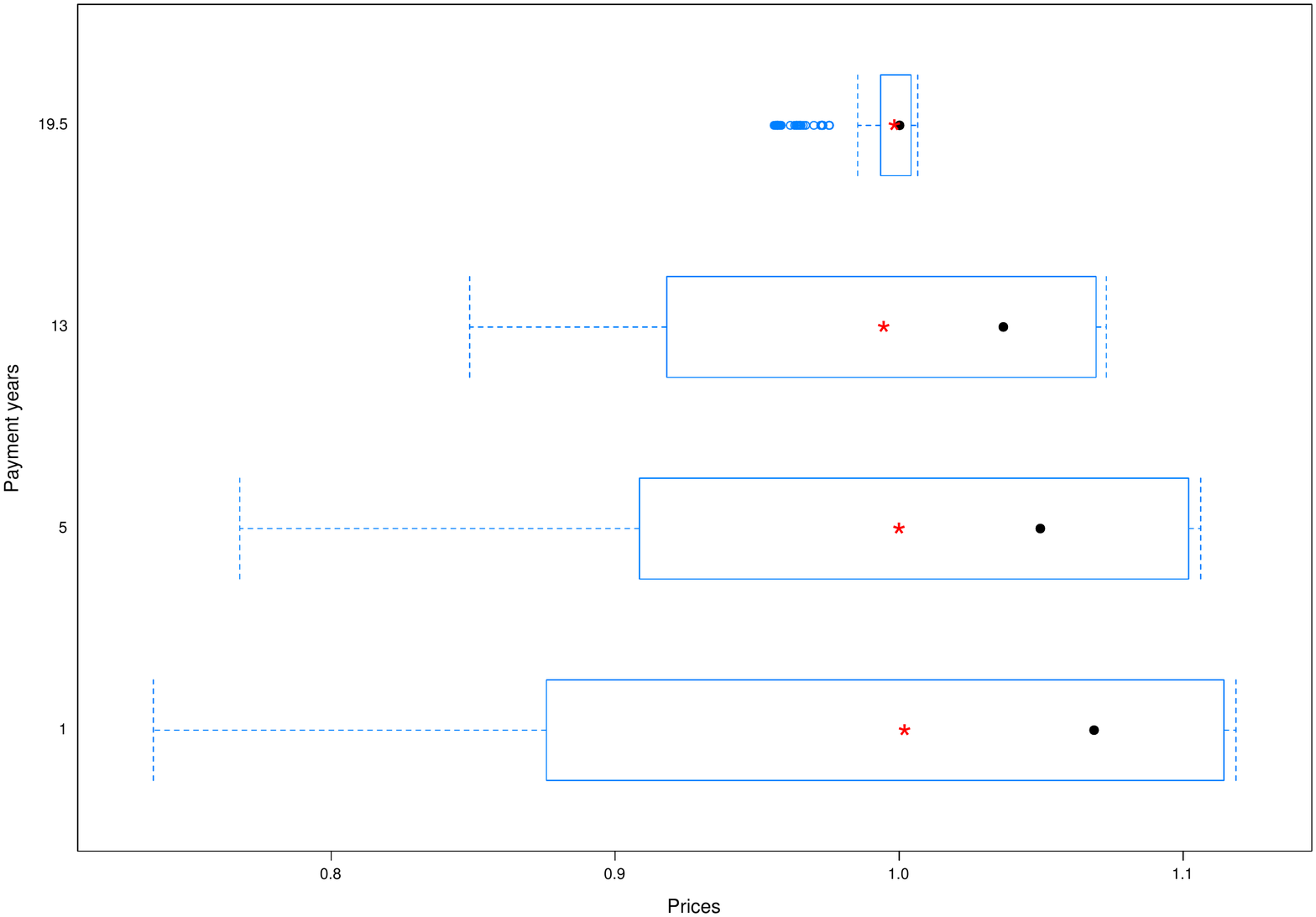}}
\caption{Price distribution of 20-yr S-CoCo with threshold 200 at 1, 5, 13 and 19.5 years.}\label{fig:BoxWiskers}
\end{figure}

However, the objective is not to provide an alternative pricing method, but to determine the future distribution of prices for risk management. We apply LSM to price a 20-yr S-CoCo for Greece, Italy and Germany, and obtain price distributions at 1, 5, 13 and 19.5  years. CDS spread and short rate dynamics are calibrated on the same set of data as in Section~\ref{sec:CoCoPricing}. Experiments are carried out for 100 regime scenarios, and 1000 CDS spread scenarios for each regime scenario. Box-Whiskers plots illustrate in Figure~\ref{fig:BoxWiskers} the distributions for $\bar{s}=200$, where a red star indicates the average and a black dot the median of the price distributions. Price distributions converge to an expected price of par at maturity and this pull-to-par phenomenon shrinks the variability of price distributions near maturity. The distributions are skewed and bimodal (bi-modality is not seen from the Box-Whiskers plot but is evident when plotting the histogram).  These results are intuitive and the contribution of our paper is to quantify them. These price distributions can be used to compute holding period returns at different horizons for risk management \citep{MulZen:1994}. In \cite{ConZen:2015b} we use holding period return distributions to illustrate how S-CoCo could improve the risk profile of a eurozone crisis country.

\clearpage

\subsection{The effect of regime switching on state contingent prices} 

To gain further insights in the performance of S-CoCo, we numerically test the effects of regime switching. Italy is used in all experiments, with thresholds 200 and 500. In the former case the standstill is activated and the results are qualitatively similar to what one would expect for Greece as well. In the later case the standstill is very rarely triggered and the results are very different from those of Greece. Results are reported again for a 20-yr S-CoCo price distribution at 1, 5, 13 and 19.5 years, but under different scenario test beds with and without regime switching. In particular:
\begin{description}
    \item[\texttt{R-OFF}] No regime switching, with the parameters used to calibrate the CDS spread model set to their historical average and simulating 5000 CDS spread and interest rate scenarios.
    \item[\texttt{R-1}] Only one scenario of regime switching between the identified regimes with 5000 CDS spread and interest rate scenarios. 
    \item[\texttt{R-100}] 100 simulations of regime switching between the identified regimes and 1000 scenarios of CDS spread and interest rates for each regime scenario.
\end{description}
 
\begin{figure} 
\subfloat[Threshold 200]{\includegraphics[width=\textwidth,height=0.6\textwidth,clip]{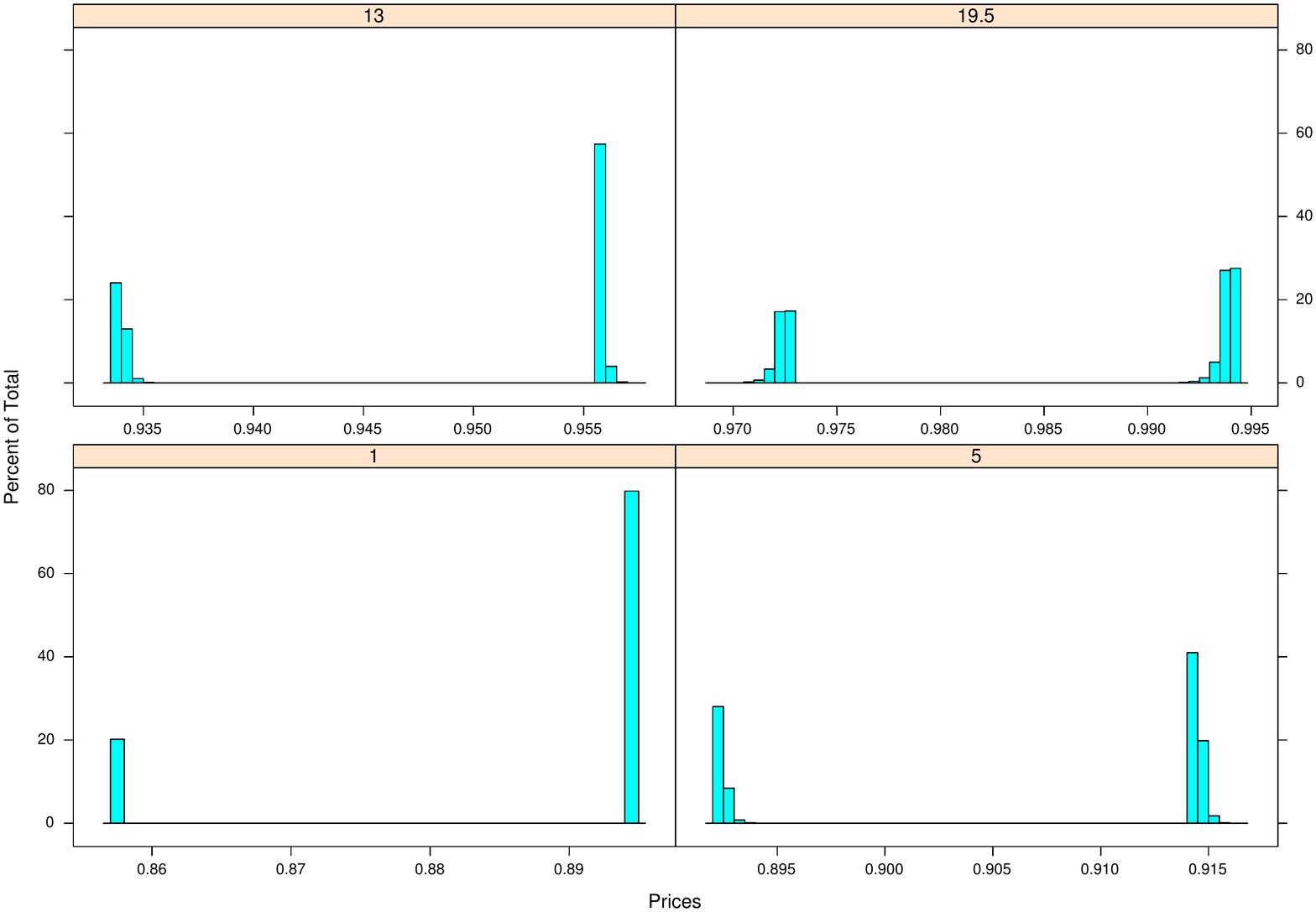}} \\
\subfloat[Threshold 500]{\includegraphics[width=\textwidth,height=0.6\textwidth,clip]{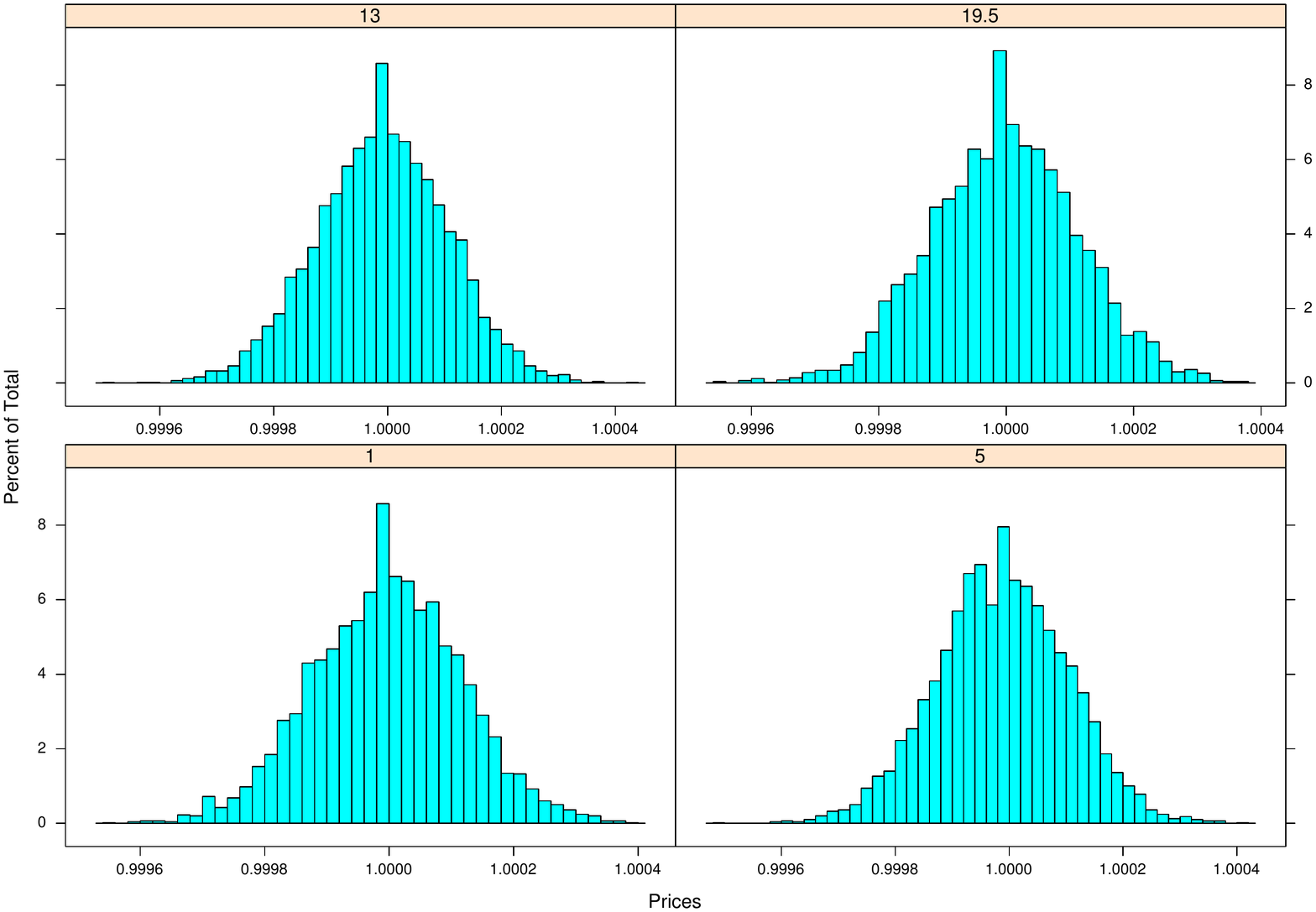}}
\caption{Price distribution of 20-yr Italian S-CoCo at different risk horizons without regime switching (test bed \texttt{R-OFF}). }\label{fig:Histogram-R-OFF}
\end{figure}

\begin{figure} 
\subfloat[Threshold 200]{\includegraphics[width=\textwidth,height=0.6\textwidth,clip]{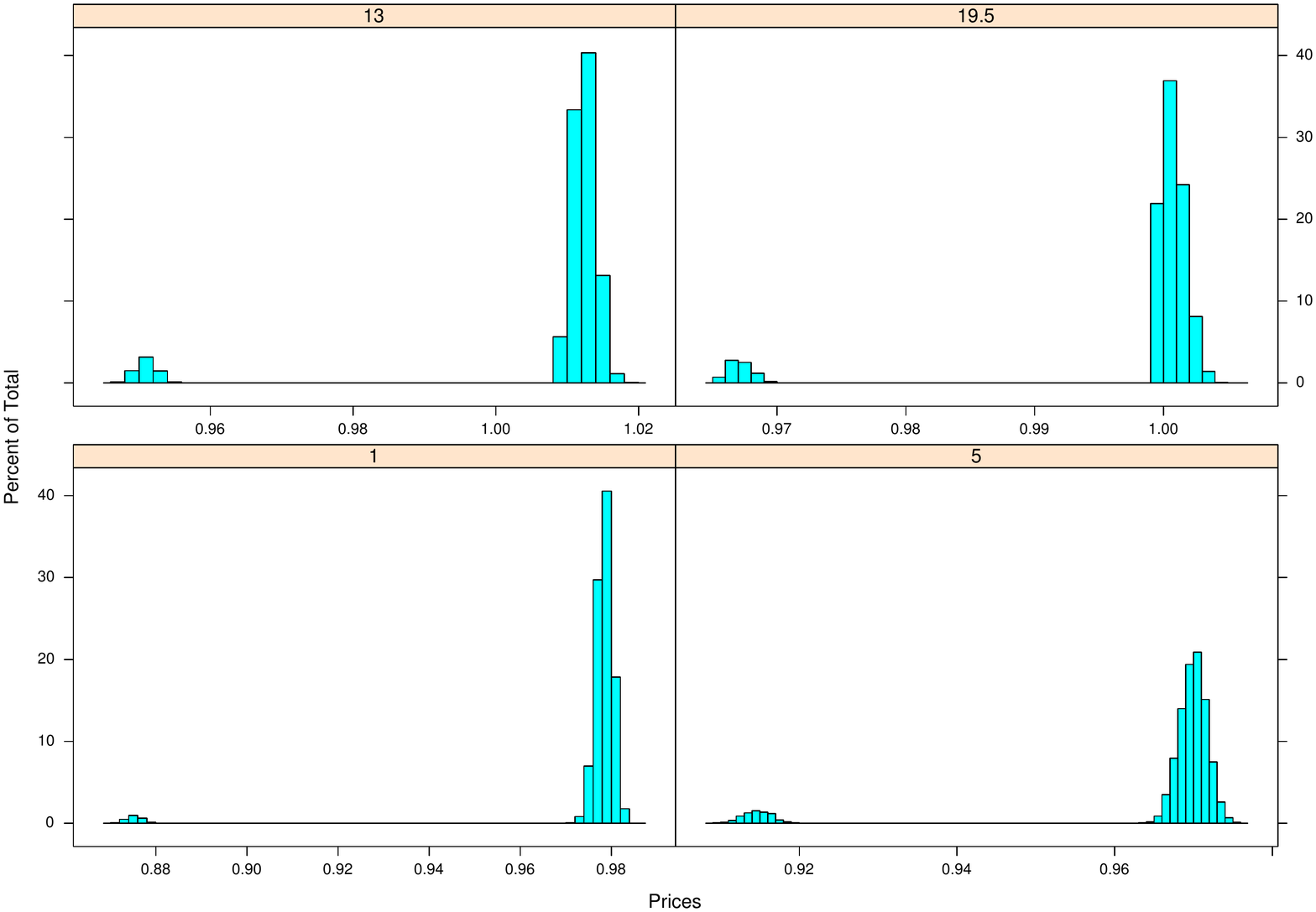}} \\
\subfloat[Threshold 500]{\includegraphics[width=\textwidth,height=0.6\textwidth,clip]{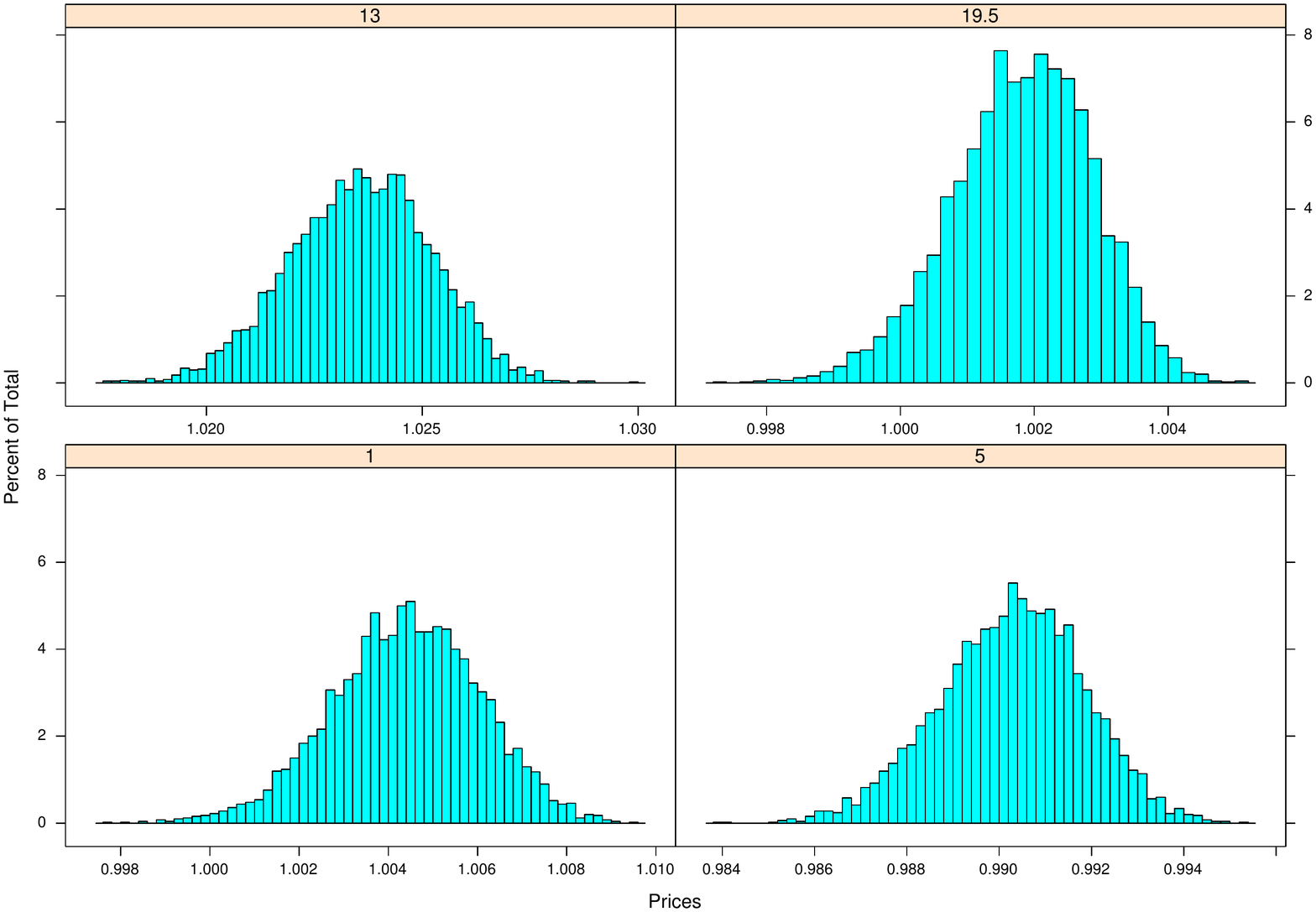}}
\caption{Price distribution of 20-yr Italian S-CoCo at different risk horizons with only one regime switching scenario (test bed \texttt{R-1}).}\label{fig:Histogram-R-1}
\end{figure}

\begin{figure} 
\subfloat[Threshold 200]{\includegraphics[width=\textwidth,height=0.6\textwidth,clip]{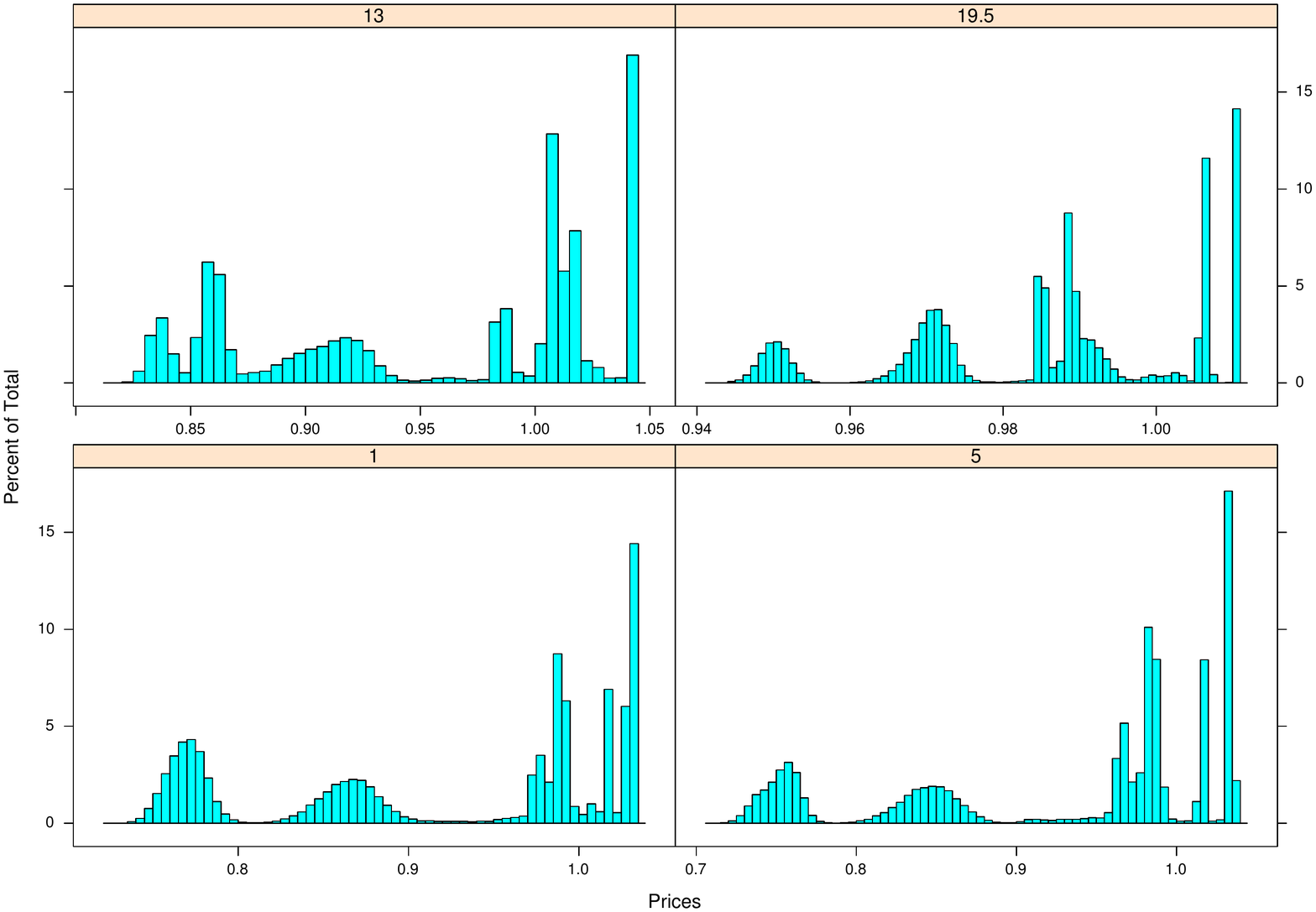}} \\
\subfloat[Threshold 500]{\includegraphics[width=\textwidth,height=0.6\textwidth,clip]{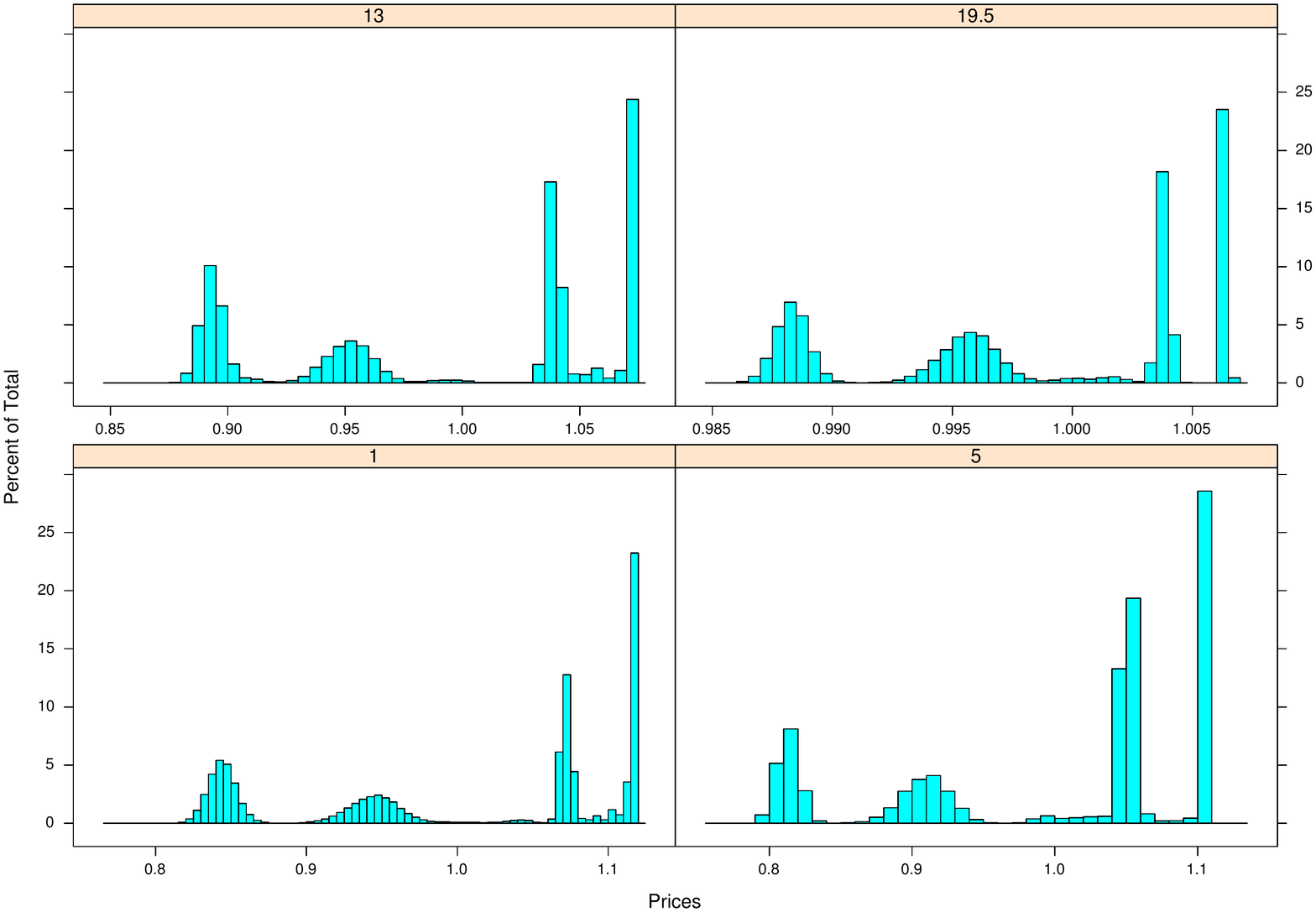}}
\caption{Price distribution of 20-yr Italian S-CoCo at different risk horizons with multiple regime switching scenarios (test bed \texttt{R-100}).}\label{fig:Histogram-R-100}
\end{figure}

Figures~\ref{fig:Histogram-R-OFF}--\ref{fig:Histogram-R-100} show the distribution of the prices for the three scenario test beds. The following observations can be made: 
\begin{enumerate}
    \item With the regime scenario simulation switched off and the CDS spread calibrated to the historical average, the S-CoCo with threshold 200 exhibits an (almost) binary distribution, while at threshold 500 its prices are just like a straight bond. Under the historical average regime the Italian CDS spreads do not exhibit sufficient variability to trigger the S-CoCo. Payment standstill becomes an extremely rare event, but with big impact.
    \item When introducing even one regime scenario, capturing the observation of the recent past that Italy may move from a tranquil regime into turbulence and even a crisis, then the distribution of prices at threshold 200 exhibits more variability. There is also a non-trivial effect for threshold 500, although significantly lower than at the 200 threshold.  
    \item Finally, when simulating properly both regime switching and CDS spreads we obtain multi-modal distributions at the risk horizon. These modalities result from a combination of regime switching and standstill triggers. 
\end{enumerate}

The multi-modality of the distributions, when simulated properly, may be disconcerting. This is inescapable when modeling events with large impact ---such as regime switching--- and limited historical data to calibrate. If we could offer a criticism to our modeling approach is that a regime derived from expert opinion ---such as ``after the Brexit referendum, Italian CDS spreads will reach levels seen at the peak of the Eurozone crisis and stay there until the Brexit issue is resolved''--- maybe more appropriate than a statistical model. If an expert opinion regime is available, the pricing model applies with {\tt R-OFF}. 

\clearpage
\subsection{Dual trigger pricing}

\cite{McDonald:2013bu} argues that bank CoCo should not be converted for idiosyncratic problems but only when the entity's difficulties come with market-wide problems. He illustrates dual trigger structures with a simple pricing example. The arguments for dual triggers seem well accepted for corporate debt where firms should be allowed to fail when they face difficulties in a benign market, but not when they face problems during a market-wide crisis. Sovereigns, on the other hand, do not fail, and \cite{ConZen:2015b} ``do not see any arguments in favor of a dual trigger, although an additional market-specific indicator could be introduced to allow for potential [sovereign] default''. However the debate on sovereign contingent debt is at an early stage, and should dual price trigger be considered necessary we could use a systemic market index such as the CBOE volatility index VIX, or the emerging markets EMBI index, or, for eurozone countries, the CDS spreads on AAA-rated sovereigns. If a sovereign's CDS threshold is breached during a systemic crisis as indicated by the market index, then the payment standstill is triggered, but for idiosyncratic crises there would be a different treatment.  

We develop here the S-CoCo pricing model with a dual trigger. The model of  Section~\ref{sec:CoCoPricing} is extended by augmenting the stochastic process  $\xi$ with the market index $v_t$,  $\xi=\left\{r_t, s_t, v_t\right\}$. (We assume that the stochastic components of $\xi$ are uncorrelated. More complex patterns of correlation between  the market index $v_t$ and $r_t$ and  $s_t$ can be introduced, but they are beyond  the scope of the present paper.) A trigger threshold $\bar{v}$ relates to the market index $v_t$, and $\mathcal{T}^l_m$, $m=1,2,\ldots, M$, denotes those time sets in which, conditioned on scenario $l$, there is a coupon payment standstill for $K_1$  periods.

For the standstill to be triggered both $s^l_t$ and $v_t^l$ must breach their respective thresholds $\bar{s}$ and $\bar{v}$ at time $t$. But we also need to model situations where the market and the country specific indices decouple, to account for idiosyncratic crises. There are different patterns of aid that can be envisioned for such eventualities, which are represented using a different standstill period, $K_2$. We embrace the view that countries in financial distress for purely idiosyncratic reasons need more help and therefore $K_2 > K_1$. 

If at time $t$, and under scenario $l$, the CDS rate $s^l_{t}$ hits $\bar{s}$, and the market index $v_t^l$ is below $\bar{v}$, coupon payments are suspended for $K_2$ periods. Denote by 
$\mathcal{V}^l_q = \left \{ t, t+1, \ldots, t+K_2 \right\}$, $q=1,2,\ldots, Q$, such time sets with the same properties as $\mathcal{T}^l_m$. The time sets defining the dual trigger mechanism are then given by 
\begin{equation}
\Lambda^l = \bigcup_{m=1}^M \mathcal{T}^l_m, \quad \Upsilon^l =\bigcup_{q=1}^Q \mathcal{V}^l_q,
\end{equation}
and the new indicator function $\mathbbm{1}_{\Upsilon^l}: \mathcal{T} \rightarrow \lbrace 0,1 \rbrace$ is
\begin{equation}
\mathbbm{1}_{\Upsilon^l}(t) = 
\left\lbrace
\begin{array}{ll}
0,&\mbox{if}\quad t \in \Upsilon^l\\
1,&\mbox{if}\quad t \notin \Upsilon^l.
\end{array}
\right. 
\end{equation}

The S-CoCo price function (\ref{eq:cocoPricing}) with a dual trigger becomes 
\begin{equation}\label{eq:cocoPricingDualTrigger}
P_0 = \frac{1}{N}\sum_{l \in \Omega} \sum_{t \in \mathcal{T}} B^l(0,t) \left(\mathbbm{1}_{\Lambda^l}(t) \cdot \mathbbm{1}_{\Upsilon^l}(t) \right) \; c + B^l(0,T+\Delta T^l).
\end{equation}
Note that, since $\mathcal{T}^l_m \cap \mathcal{V}^l_q = \emptyset$, for any $m=1,2,\ldots,M,$ and $q=1,2,\ldots,Q$, then also $\Lambda^l \cap \Upsilon^l = \emptyset$, and the product of the two indicator functions correctly represents the dual trigger mechanism.

\section{Conclusions} \label{s:conclusions}

We developed a pricing model for sovereign contingent convertible bonds with payment standstill that captures the regime-switching nature of the triggering process. We adapt an existing single-factor tractable stochastic model of spread-returns with mean-reversion to model spread levels converging to a long-term steady state value estimated from market data, whereby the steady state is modeled by a novel regime switching Markov process model. The Monte Carlo simulation pricing model is embedded in the Longstaff-Schwartz framework to compute state contingent prices at some risk horizon. This facilitates risk management. 

Extensive numerical experiments illustrate the performance of the models and shed light on the performance of sovereign contingent debt. In particular, we observe the skewed distribution of prices at the risk horizon, the pull-to-par phenomenon as securities approach maturity, and the multi-modality of the price distribution as the underlying CDS process switches regimes and/or the payment standstill is triggered.

The models are applied to S-CoCo designs for Greece, Italy, and Germany, in order to illustrate how these instruments would be priced for countries under different economic conditions. The results are intuitive and the contribution of the paper is in providing a model to quantify prices and holding period returns. Such a model is an essential tool if sovereign contingent debt is to receive attention and eventual acceptance as a practical financial innovation response to the problem of debt restructuring in sovereign debt crises. In a companion paper, we show how these models can be used to develop a sovereign debt risk optimization model to improve a country's risk profile \citep{ConZen:2015b}.

\appendix
\section{Appendix. Asymptotic modeling of the scenario generating process} \label{app:asymptotic}

We determine the parameters of the model for CDS spread return to identify its asymptotic dynamics. We start from the discrete time model of \citep[cf. eqn.~(2), or eqn.~(6) without the jump term]{Donoghue:2014}  and derive a set of conditions on the asymptotic moments to be matched with empirically estimated values. To simplify the notation in their eqns.~(2) or (6), set $k_0 = \gamma$, $k_1= \alpha + \beta$ and $k_2= \alpha \beta$, to get 
\begin{equation}
\Delta r_t = \left(k_0 -k_1 r_t - k_2 \sum_{s=0}^t r_{s} \Delta t \right) \Delta t + \sigma w_t,\label{eq:StochasticSpreadReturnDynamics}
\end{equation}
where $r_t$ is the return at time $t$ and $w_t \sim \mathcal{N}(0, \Delta t)$.

The simulation model is made up of two stochastic equations, one for the CDS and one for the interest rate, identical structure given by (\ref{eq:StochasticSpreadReturnDynamics}). In case the two factors are correlated, we need two noise components, $\epsilon^1_t,\epsilon^2_t \sim \mathcal{N}(0, \Delta t)$, with  $\rho\left(\epsilon^1_t,\epsilon^2_t\right)=0$. It can be easily shown that the two processes $w^1_t$ and $w^2_t$, given by
\begin{align}
    w^1_t &= \epsilon^1_t\\
    w^2_t &= \rho \epsilon^1_t + \epsilon^2_t\sqrt{(1-\rho^2)},
\end{align}
have correlation $\rho$, to be estimated from available historical time series.

Following \cite{Donoghue:2014}, for $t \rightarrow \infty$, 
we  have
\begin{align}
\mathbb{E}[r_t] &= 0\\
\mbox{var}[r_t] & = \frac{\sigma^2}{2 k_1}\label{eq:AsymptoticReturnSpreadVariance}\\ 
\mathbb{E}[C_t] &= \frac{k_0}{k_2}\\
\mbox{var}[C_t] &= \frac{\sigma^2}{2 k_1 k_2},
\end{align}
where $C_t = \sum_{s=0}^t r_s \Delta t$ and $C_t$ is normally distributed.

The spread process $S_t = S_0 \exp(C_t)$ is log-normally distributed with 
\begin{align}
\mathbb{E}[S_t] &= S_0 \exp\left( \frac{k_0}{k_2} + \frac{\sigma^2}{4 k_1 k_2}\right)\label{eq:AsymptoticSpread}\\
\mbox{var}[S_t] &= S_0^2 \exp\left( 2 \frac{k_0}{k_2} + \frac{\sigma^2}{2 k_1 k_2}\right) \left[ \exp\left( \frac{\sigma^2}{2k_1 k_2}\right)-1\right]\label{eq:AsymptoticSpreadVariance}.
\end{align}

We now have three equations in the four unknowns of the stochastic dynamics (\ref{eq:StochasticSpreadReturnDynamics}). We need one additional condition which we derive from the squared changes $\mathbb{E}[(\Delta r_t)^2]$, which is a measure of the smoothness of the process. With some standard assumptions for stochastic processes, namely that  $\mathbb{E}[w_t C_t] = 0$, $\mathbb{E}[r_t C_t] = 0$ and  $\mathbb{E}[r_t w_t] = 0$, and using simple algebra, we obtain
\begin{equation}
\mathbb{E}[(\Delta r_t)^2] = \frac{\sigma^2}{2} \left( k_1 + \frac{k_2}{k_1} + 2\right)\label{eq:AsymptoticSmoothness}.
\end{equation}
A sample estimate $\hat{s}^2$ for $\mathbb{E}[(\Delta r_t)^2]$ is given by 
\begin{equation}\label{app:EstimatedSmoothedParameter}
\hat{s}^2 = \frac{1}{N} \sum_{t=1}^N \left(r_t - r_{t-1} \right)^2.
\end{equation}

The theoretical moments defined by (\ref{eq:AsymptoticReturnSpreadVariance}), (\ref{eq:AsymptoticSpread}), (\ref{eq:AsymptoticSpreadVariance}),  and of the smoothness  (\ref{eq:AsymptoticSmoothness}) are then matched to the empirical observations. We denote by $\hat{S}$ the asymptotic CDS spread level, by $\hat{\sigma}_S$ the asymptotic variance of CDS spread level, by $\hat{\sigma}_r$ the asymptotic variance of CDS spread returns, and by $\hat{s}^2$ the smoothness of the CDS spread level.
(These quantities are estimated for each regime separately if regime switching is manifested in the empirical data, e.g., Table~\ref{tab:meanCDS}.) Denoting by $T_{\tau}$ the set of time periods in regime $\tau$, we obtain the following moment estimates for the regime:

\begin{align}
\hat{S} &= \frac{1}{|T_{\tau}|} \sum_{t \in T_{\tau}} S_t\label{eq:EstAsymptoticSpread}\\
\hat{\sigma}^2_S &= \frac{1}{|T_{\tau}|} \sum_{t \in T_{\tau}} \left( S_t - \hat{S} \right)^2 \label{eq:EstAsymptoticSpreadVariance}\\
\hat{\sigma}^2_r &= \frac{1}{|T_{\tau}|} \sum_{t \in T_{\tau}} \left( r_t - \hat{r} \right)^2. \label{eq:EstAsymptoticReturnSpreadVariance}
\end{align}
Similarly, we have the estimate of the smoothness of the regime:
\begin{equation}
\hat{s}^2 = \frac{1}{|T_{\tau}|} \sum_{t \in T_{\tau}} \left(r_t - r_{t-1} \right)^2. \label{eq:EstSmoothness}
\end{equation}

If $S_0$ denotes the starting value of the CDS spread for the selected regime, we can match the theoretical moments to their estimated values solving the system of nonlinear equations in $k_0$, $k_1$, $k_2$ and $\sigma$:
\begin{align}
   \exp\left( \frac{k_0}{k_2} + \frac{\sigma^2}{4 k_1 k_2}\right) &= \frac{\hat{S}}{S_0}\label{eq:FirstOfSystem}\\
   \exp\left( 2 \frac{k_0}{k_2} + \frac{\sigma^2}{2 k_1 k_2}\right) \left[ \exp\left( \frac{\sigma^2}{2k_1 k_2}\right)-1\right] &= \frac{\hat{\sigma}^2_S}{S_0^2}\\
\frac{\sigma^2}{2 k_1} &= \hat{\sigma}^2_r\\
\frac{\sigma^2}{2} \left( k_1 + \frac{k_2}{k_1} + 2\right) &= \hat{s}^2\label{eq:LastOfSystem}.
\end{align}
(Of course the right hand side parameters do not have to be estimated from historical data, but can be values assumed or estimated by the user. For instance, the user may wish to price the instruments for extreme values of the moments, or the values implied by the rating of a country.)

The closed form solution to the system of equations (\ref{eq:FirstOfSystem})--(\ref{eq:LastOfSystem}) is given by
\begin{align}
k_0 &= \frac{\hat{\sigma}^2_r}{\log\left( 1+ \frac{\hat{\sigma}^2_S}{\hat{S}^2}\right)} \log\left( \frac{\hat{S}}{S_0}\right) - \frac{1}{2}\hat{\sigma}^2_r\\
k_1 &= \frac{\sigma^2}{2 \hat{\sigma}^2_r}\\
k_2 &= \frac{\hat{\sigma}^2_r}{\log\left( 1+ \frac{\hat{\sigma}^2_S}{\hat{S}^2}\right)}\\
\sigma^2 &= 2 \hat{\sigma}^2_r \left[ -1 + \sqrt{1 + \frac{\hat{s}^2 - k_2 \hat{\sigma}^2_r}{4\hat{\sigma}^2_r}} \, \right].
\end{align}
Finally, to ensure that $\sigma \in \re^+$ we need  $\hat{s}^2 - k_2 \hat{\sigma}^2_r > 0$.

\begin{figure}
\includegraphics[width=\textwidth,height=0.5\textwidth,clip]{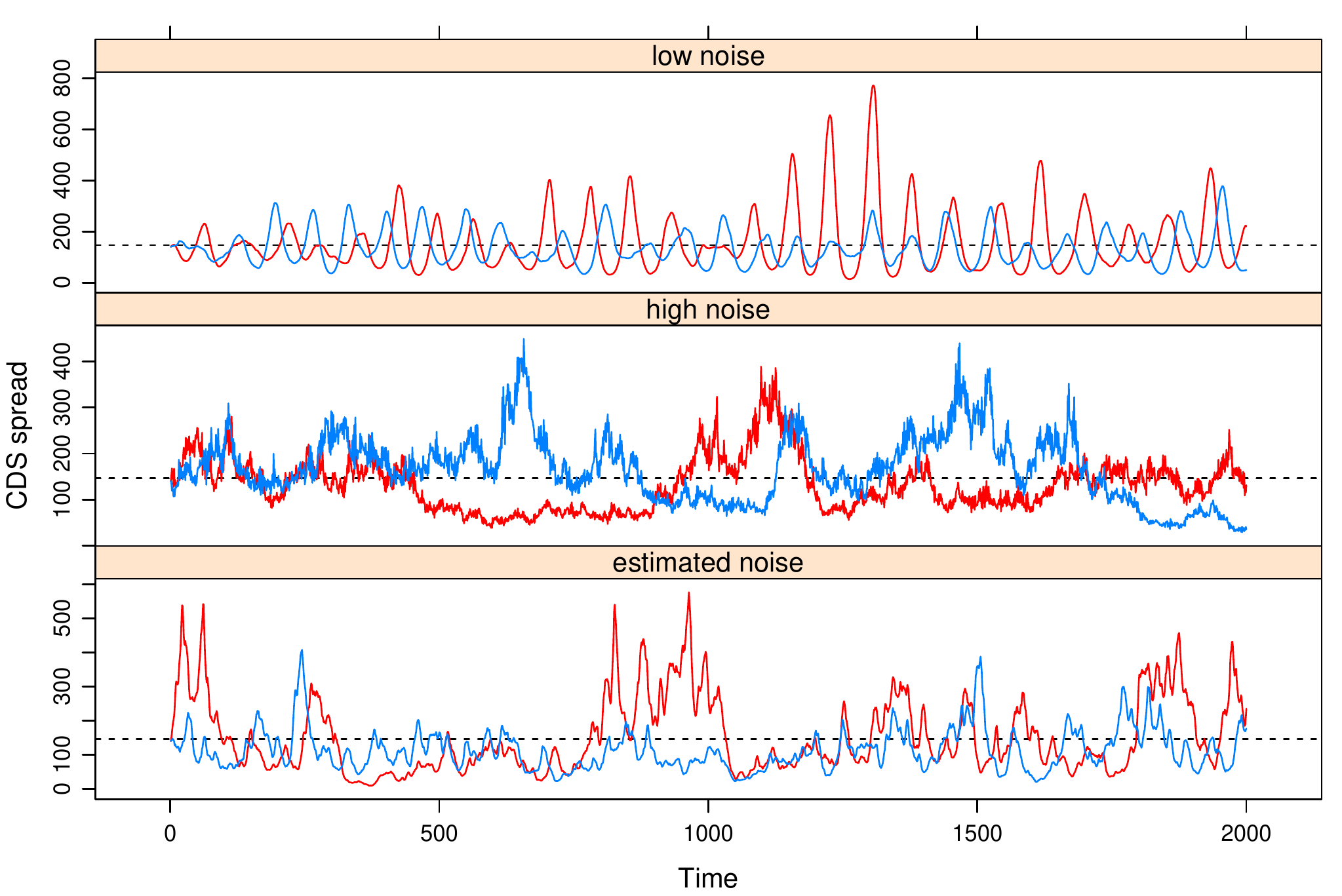}
\caption{The effect of the parameter $\sigma$ on the smoothness of CDS spread dynamics. In the upper panel we use a low value of the parameter $\sigma$, in the middle panel we use a high value of $\sigma$, and in the bottom panel we use $\sigma$ estimated from eqn.~(\ref{app:EstimatedSmoothedParameter}). The dotted lines show the asymptotic level of the CDS spread.}\label{fig:NoiseEffect}
\end{figure}

We point out the role of the noise term $\sigma$ on the smoothness of the process. Figure~\ref{fig:NoiseEffect} illustrates the effect of $\sigma$ on two paths generated obtained from (\ref{eq:StochasticSpreadReturnDynamics}). Observe that the lower the $\sigma$, the smoother is the generated curve, so $\sigma$  relates to $\mathbb{E}[(\Delta r_t)^2]$. In the same figure we illustrate the two paths when calibrated on a value estimated from historical data using the system of equations above.

\newpage

\bibliographystyle{plainnat} 

\end{document}